\documentclass[fleqn,twoside]{article}
\usepackage{amsmath,amssymb,espcrc2,color,cite}
\usepackage[usenames,dvipsnames]{xcolor}
\usepackage[percent]{overpic}

\usepackage{graphicx,soul}
\usepackage[figuresright]{rotating}
\usepackage{ulem}
\usepackage{capt-of}

\usepackage{hyperref}
\hypersetup{
    colorlinks=true,
    linkcolor=blue,
    filecolor=magenta,      
    urlcolor=cyan,
    citecolor=blue,
    allcolors=blue
}

\newcommand{\be}{\begin{equation}}
\newcommand{\ee}{\end{equation}}
\newcommand{\bea}{\begin{eqnarray}}
\newcommand{\eea}{\end{eqnarray}}

\definecolor{orange}{rgb}{1,0.5,0}

\definecolor{Gray}{named}{Gray}

\graphicspath{{figs/}}

\title{\vspace{-4cm}
\bf \LARGE Prospects of searches for 
long-lived charged particles with MoEDAL \\ \vspace{0.5em}}

\author{
{\bf B.S.~Acharya}\address[trieste]{The Abdus Salam International Centre for Theoretical Physics, Strada Costiera 11, Trieste, Italy}$^,$\address[kcl]{Theoretical Particle Physics and Cosmology Group, Department of Physics, King's College London, Strand, London WC2R~2LS, UK},
{\bf A.~De~Roeck}\address[cern-ex]{Experimental Physics Department, CERN, CH–1211 Geneva 23, Switzerland}$^,$\address[antwerp]{Antwerp University, B–2610 Wilrijk, Belgium},
{\bf J.~Ellis$^{\rm b}$}$^,$\address[cern-th]{Theoretical Physics Department, CERN, CH–1211 Geneva 23, Switzerland}$^,$\address[Tallinn]{National Institute of Chemical Physics \& Biophysics, R\"avala 10, 10143 Tallinn, Estonia},
{\bf D.K.~Ghosh}\address[iacs]{School of Physical Sciences, Indian Association for the Cultivation of Science, 2A \& 2B Raja S.C.~Mullick Road, Kolkata 700 032, India},
{\bf R.~Mase\l{}ek}\address[warsaw]{Institute of Theoretical Physics, Faculty of Physics, University of Warsaw, ul.~Pasteura 5, PL--02--093 Warsaw, Poland},
{\bf G.~Panizzo}$^{\rm a,}$\address[inst2]{INFN Gruppo Collegato di Udine, Sezione di Trieste, Udine},
{\bf J.L.~Pinfold}\address[alberta]{Physics Department, University of Alberta, Edmonton, Alberta T6G~2E4, Canada},
{\bf K.~Sakurai$^{\rm h}$},
{\bf A. Shaa}$^{\rm h}$,
{\bf A.~Wall}\address[alabama]{University of Alabama, Department of Physics, Tuscaloosa, Alabama, USA}
}

\begin{document}
\begin{abstract}
\vspace{0.5cm}
\begin{center}
{\bf {\large Abstract}}
\end{center}
\vspace{0.2cm}
We study the prospects of searches for exotic long-lived 
particles with the MoEDAL detector at the LHC, 
assuming the integrated luminosity of 30 fb$^{-1}$ 
that is expected at the end of Run~3. MoEDAL incorporates
nuclear track detectors deployed 
a few metres away from the interaction point, which
are sensitive to any highly-ionizing particles.
Hence MoEDAL is able to detect 
singly- or doubly-charged particles 
{with low velocities $\beta < 0.15$ or $< 0.3$, respectively, and}
lifetimes larger than ${\cal O}(1) \,{\rm m}/c$.
{We examine the MoEDAL sensitivity to 
various singly-charged supersymmetric particles with long lifetimes
and to several types of doubly-charged long-lived particles with 
different spins and SU(2) charges. We compare the prospective MoEDAL
mass reaches to current limits from ATLAS and CMS, which involve
auxiliary analysis assumptions. MoEDAL searches for doubly-charged
fermions are particularly competitive.}

\vspace{1cm}
\begin{center}
{\tt KCL-PH-TH/2020-18, CERN-TH-2020-060}
\end{center}
\vspace{4cm}
\end{abstract}
\thispagestyle{empty}
\newpage


\maketitle


\newpage

\section{Introduction}
\label{sec:intro}

One of the key priorities of the LHC experimental programme is to
search for new particles beyond the Standard Model, or at least set
robust constraints on their possible existence. Priorities
during Runs 1 and 2 of the LHC included scans for bumps in invariant-mass
spectra, searches for excesses of missing transverse energy in events with 
various topologies, and precision tests of Standard Model (SM) predictions.
With the striking exception of the discovery of the Higgs boson, none of
these searches has yet borne any fruit.

In parallel with these mainstream searches for new particles, there has
been growing interest in less-orthodox searches. These have included
searches for long-lived particles (LLPs),
which we review in Section~\ref{sec:previous}. These long-lived particles may arise in various scenarios beyond the SM 
that satisfy any of the following conditions: {(i)}
small couplings, (ii) nearly degenerate masses 
or (iii) heavy intermediate (virtual) particles mediating decays\cite{Lee:2018pag,Alimena:2019zri,Curtin:2018mvb}.
We expect such searches to receive increased attention during LHC Run~3.

{During Runs 1 and 2 of the LHC the search for magnetically-charged particles was the main objective
of the MoEDAL experiment, which has published
{limits} from searches for magnetic monopoles~\cite{MoEDALmono} and dyons~\cite{MoEDALdyon}}. The ATLAS Collaboration has also published
limits on magnetic monopole production~\cite{ATLASmono}.

As we discuss in Section~\ref{sec:moedal}
one of the key features of the MoEDAL
detector~\cite{Acharya:2014nyr} is its array of
panels of nuclear track detectors (NTDs), which can {register} anomalously
heavy ionization as would be produced by singly-charged particles with
velocities $\beta \equiv v/c < 0.15$, or by doubly-charged particles {with $\beta < 0.3$}.
The MoEDAL NTD panels are located at distances $\sim 2$\,m from the
interaction point, and are therefore sensitive to long-lived heavily-ionizing
particles with lifetimes $\tau$ longer than ${\cal O}(1)$\,m/$c$ \cite{Felea:2020cvf, Sakurai:2019bac}.
{We discuss in Section~\ref{sec:framework} the analysis framework that
we use to study long-lived charged particles, which we apply
to a couple of well-motivated theoretic models} in Sections~\ref{sec:susy} and \ref{sec:dcp}. 

One such model is supersymmetry,
which has been the object of many {other} searches at the LHC~\cite{ATLAStwiki,CMStwiki}. The minimal
supersymmetric extension of the SM (the MSSM) does not contain any doubly-charged
particles, but many of its singly-charged particles are candidates to be long-lived,
and several mechanisms suggest this possibility \cite{Martin:1997ns,Haber:1984rc,rohini_book,Baer:2006rs }. The lightest supersymmetric particle (LSP) would be stable if R-parity is conserved, 
but could be unstable {with a long lifetime}
if R-parity is weakly broken \cite{Barbier:2004ez,Barry:2013nva}. Even if R-parity is conserved, the next-to-lightest supersymmetric
particle (NLSP) could be long-lived if it is very nearly degenerate with the LSP, as in
coannihilation scenarios, or if the supersymmetric particle(s) mediating its decay is (are)
very heavy, as in scenarios with split supersymmetry \cite{ArkaniHamed:2004fb, Giudice:2004tc}.
Another possibility is that the LSP has only very weak interactions, e.g., if it is the gravitino or some other particle
with gravitational-strength interactions, or if it is located in a well-sequestered
hidden sector. In either case, the lightest supersymmetric particle in the visible
sector could be long-lived \cite{Randall:1998uk,Giudice:1998xp,Giudice:1998bp}. 

Prospects for discovering long-lived sparticles with MoEDAL was  first discussed 
in~\cite{Felea:2020cvf, Sakurai:2019bac}.  These studies considered 
a double long-lived (LL) cascade chain: $pp \to \tilde g \tilde g$, $\tilde g \to jj [\tilde \chi_1^0]_{\rm LL}$, $[\tilde \chi_1^0]_{\rm LL} \to \tau [\tilde \tau]_{\rm LL}$, 
assuming that the $\tilde \chi_1^0$ and $\tilde \tau$ are both long-lived.
In this paper we extend this discussion to consider direct pair production of general meta-stable sparticles,
including
charged {R-hadrons containing a gluino, squark or stop},
as well as Winos, Higgsinos and sleptons, without making any specific assumptions about the
models in which they appear \cite{Farrar:1978xj}.

{A} second scenario that may lead to a long-lived heavy fermion is a Type-III seesaw model \cite{Foot:1988aq,Bajc:2006ia,Arhrib:2009mz}, in which the SM is augmented with at least two $SU(2)_L$-triplet fermion fields $(\Sigma)$ with $Y=0$. The observed neutrino mass is given by $m_\nu \approx Y^2_\nu v^2/m_{\Sigma}$, where $Y_\nu$ is the Dirac Yukawa coupling, $v$ is the SM vev and $M_\Sigma$ is the mass of the heavy triplet fermion. In this model, {radiative corrections generate a mass splitting $m_{\Sigma^\pm} - m_{\Sigma^0} > m_\pi^\pm$~\cite{Simple,Jana:2019tdm} so that
the decay $\Sigma^\pm \to \Sigma^0 + \pi^\pm$ is kinematically allowed, with a lifetime ${\cal O}(10^{-8})$~s that leads to very soft charged pion but no long-lived charged-particle track. However, if there are additional contributions to the $\Sigma^{\pm, 0}$ masses that reduce the mass difference so that the two-body decay is not allowed, the dominant decay of the $\Sigma^\pm$ would be three-body with a lifetime ${\cal O}(10^{-6})$~s. The latter case would lead to a detectable long-lived charged particle signature similar to that of the Wino, as we discuss later in the paper.}

A long-lived doubly-charged scalar particle may appear in a number of models. One example is a 
Type-II seesaw model of neutrino masses, in which the SM is supplemented by a complex $SU(2)_L$ triplet of 
scalar fields with hypercharge $Y = 2$ \cite{Schechter:1980gr, Magg:1980ut, Cheng:1980qt,Lazarides:1980nt, Mohapatra:1980yp, Lindner:2016bgg,Melfo:2011nx,Dev:2013ff,Ghosh:2017pxl}, and other scenarios for doubly-charged scalars are mentioned later.
Another possibility for a doubly-charged particle is a spin-1/2 particle, such as a
doubly-charged Higgsino as appears in supersymmetric L-R 
models~\cite{Kuchimanchi:1993jg,Babu:2008ep,Francis:1990pi,Huitu:1993uv,Frank:2014kma}.
{One can also simply add doubly-charged scalars and fermions in 
  various $SU(2)_L$ representations to the particle content of the SM, then write down the SM gauge-invariant interaction terms involving these new fields and study their phenomenology~\cite{Delgado:2011iz,Alloul:2013raa}.}
{We consider here doubly-charged particles that may be scalars or fermions and either singlets or triplets of the SM $SU(2)_L$ gauge group, without making any specific assumptions about the models in which they appear.}~\footnote{Detailed  discussions on various theoretical models of long-lived stable particles and their signatures can be found in\cite{Lee:2018pag,Alimena:2019zri,Curtin:2018mvb,Dev:2018kpa, Antusch:2018svb,Banerjee:2019ktv,Bulekov:2017mkb}.}

{Both ATLAS~\cite{Aaboud:2018hdl,Aaboud:2019trc} and
CMS~\cite{Khachatryan:2016sfv} have set limits on certain species of long-lived supersymmetric particles, ATLAS has also considered multi-charged particles~\cite{Aaboud:2018kbe}, and
CMS~\cite{Khachatryan:2016sfv} has also considered one example of a doubly-charged fermion as well as multi-charged particles~\cite{Chatrchyan:2013oca}, as we review in Section~\ref{sec:previous}. However, the ATLAS and CMS searches both employed auxiliary signatures, in particular  either $E_{\rm T}^{\rm miss}$ or muon triggers \cite{Lee:2018pag,Alimena:2019zri,Curtin:2018mvb}.
Not all models containing long-lived charged particles
would yield a large $E_{\rm T}^{\rm miss}$ signature, and therefore the data sets used are a combination of 
both these complementary triggers, so as to reduce substantially
the model dependence of the search.
The MoEDAL analysis would not
invoke any such a signatures, and hence is fully model-independent. In Table~\ref{tab:susy_limit}
we compare the prospective sensitivities of MoEDAL for a wide selection of supersymmetric 
candidates with the results of ATLAS and CMS, and Table~\ref{tab:dcp_limit} shows a similar
comparison for a wide selection of doubly-charged candidates.}

{We note in passing that the parameter spaces of both the 
Type-II and-III seesaw models, as well as L-R symmetric models, are constrained by various theoretical considerations and 
low-energy experimental data~\cite{Melfo:2011nx,Dev:2013ff,Ghosh:2017pxl,Dev:2018kpa,Antusch:2018svb,Biggio:2019eeo}.  
Since our goal in this paper is to estimate the MoEDAL sensitivities on these exotic particles in a model-independent way, we refrain from discussing further these indirect model-dependent limits.}

{The outline of our paper is as follows. In Section~\ref{sec:previous} we review previous searches for long-lived charged particles at the LHC by ATLAS and CMS. Then, in Section~\ref{sec:moedal} we review relevant
aspects of the MoEDAL detector, and in Section~\ref{sec:framework} we describe the analysis
framework we use. The prospective sensitivities of MoEDAL for long-lived supersymmetric
particles are obtained in Section~\ref{sec:susy} and those for doubly-charged particles in
Section~\ref{sec:dcp}. Finally, Section~\ref{sec:conx} summarises our conclusions.}


\section{Previous LHC Searches}
\label{sec:previous}

{There have been previous searches for 
{long-lived particles (LLPs)}
at the LHC by the ATLAS~\cite{Aaboud:2018hdl,Aaboud:2019trc,Aaboud:2017iio,Aaboud:2017mpt}
and CMS~\cite{Khachatryan:2016sfv,Sirunyan:2018pwn,Sirunyan:2020pjd,Chatrchyan:2013oca} Collaborations. These searches are based on observables 
related to ionization loss, $dE/dx$, displaced tracks, displaced vertices, delayed decays and timing
information on slow{ly}-moving massive particles. Most of these unusual events are associated with conventional final states involving photons, charged leptons, jets and missing transverse energy, $E^{\rm miss}_T$. In some cases those final states occur as the decay product{s} of {LLPs}, {whilst} in other cases {they} are produced in association with the {LLPs}. In all these searches, the $p_T$ of charged leptons, photons or jets and $E^{\rm miss}_T$ were used to trigger the event selection. 

The ATLAS and CMS collaborations classified their {LLP} searches into two broad classes: (i) {heavy stable charged particles (HSCPs)} (that may decay outside the detector)~\cite{Aaboud:2018hdl,Aaboud:2019trc,Aaboud:2017iio,Khachatryan:2016sfv,Chatrchyan:2013oca} and (ii) decaying LLPs~\cite{Aaboud:2017mpt,Sirunyan:2018pwn,Sirunyan:2020pjd}. In the first case, the HSCP gives rise to signatures that are very different from traditional
 prompt signatures. {Singly-charged ($\mid Q\mid = e$) HSCPs}
 typically travel with speeds $\beta = v/c < 1$, giving rise to an ionization loss $(dE/dx)$ that is  different from that of {minimum-ionising} SM particles. Due to their slower speeds, these particles 
would also take an anomalously long time-of-flight (ToF) to reach the muon chamber. Using the information on ionization loss and ToF, the ATLAS and CMS collaborations set lower limits on the masses of such {HSCPs}, mainly in the context of supersymmetry and its variants. The ATLAS collaboration used a {dataset} of 36.1~fb$^{-1}$ {collected} during the 13 TeV runs to look for such exotic particles. No significant
signal events were observed over the expected background, {and} $95\%$ CL upper limits {were set} on the production cross-sections of long-lived R-hadrons, as well as pairs of staus and charginos. {These} upper limit{s} 
can be translated into lower limits on the masses of a long-lived gluino, sbottom, stop, stau and chargino of 2000, 1250, 1340 and 1090 GeV, respectively~\cite{Aaboud:2019trc}.~\footnote{{A shorter-lived coloured  LSP could decay inside a detector after hadronization to form an R-hadron, e.g., a gluino hadron may decay into pair of SM quark jets and the lightest neutralino, $\chi^0_1$. Using the same data set, the ATLAS Collaboration set a lower mass limit of 1290 - 2060 GeV on the gluino, assuming pair production followed by such a
decay with $m_{\chi^0_1} = 100$~GeV~\cite{Aaboud:2018hdl}.}}

{The ATLAS Collaboration also performed a 
dedicated search for the anomalous ionisation signal arising
from the pair production of $SU(3)_C$ and $SU(2)_L$ singlet 
multi-charged 
{($\mid Q\mid = Z e $, $ 2 \leq Z \leq 7$)}
HSCPs with spin-1/2
in the mass range from 50 to 1400 GeV 
 using an integrated luminosity of 
$36.1{\rm fb}^{-1}$~\cite{Aaboud:2018kbe}. Non-observation of any significant deviation from the standard model background translates into $95\%$ CL upper limits on the Drell-Yan pair production cross-section as a function of the charge of a lepton-like {HSCP} for several values of $Z$ between 2 and 7. This upper limit on the pair production cross-section can be recast into
$95\%$ CL lower limits on the masses of lepton-like HSCPs for charges $\mid Q\mid = Z e: 2 \leq Z \leq 7$, ranging between 
980 GeV and 1220 GeV for $Z=2$ and $Z=7$, respectively. 
The CMS Collaboration 
studied the pair production of $SU(3)_C$ and $SU(2)_L$ singlet singly-charged $(Z=1)$, multiply-charged $(Z > 1)$ 
and fractionally-charged $(Z< 1)$ spin-1/2 {HSCPs}
produced via the Drell-Yan process during Run~1 of the LHC at $\sqrt{s} = 7$ and 8 TeV, with
integrated luminosities of $5~{\rm fb}^{-1}$ and $18.8~{\rm fb}^{-1}$ respectively. 
The $95\%$ CL lower limit on the masses
are 480 $(Z=2/3)$, 574 $(Z=1)$, 685 $(Z=2)$, 796 $(Z=5)$, 781 $(Z=6)$, 757 $(Z=7)$ and 715 $(Z=8)$ GeV, respectively~\cite{Chatrchyan:2013oca}.}

The CMS Collaboration also put $95\%$ CL lower mass bounds on the gluino, stop and stau of 1610, 1040 and 490 GeV, respectively, from the production of HSCPs at the 13 TeV LHC run using a data sample from an integrated luminosity of 2.5~fb$^{-1}$. The corresponding mass limits on $\mid Q \mid = 1e$ $(2 e)$ lepton-like fermions of 550 (680) GeV were obtained by the CMS collaboration~\cite{Khachatryan:2016sfv}}.~\footnote{{A dedicated search by the CMS Collaboration for an anomalous signal arising from disappearing charged tracks using the data collected in Run~2 during the 13 TeV run 
of the LHC,  corresponding to an integrated luminosity of $140~{\rm fb}^{-1}$, 
set a $95\%$~CL lower 
limit $m_{\chi^{\pm}_1} = 880$ (700) GeV for a purely wino LSP in the AMSB model~\cite{ Giudice:1998xp,Randall:1998uk} for $\tau_{\chi^\pm_1} = 3$ (33) ns, respectively \cite{Sirunyan:2020pjd}.}
{The CMS Collaboration used a similar signature and $35.9$~fb$^{-1}$ of data from the 13 TeV LHC run~\cite{Sirunyan:2017qkz} to constrain the $SU(2)_L$-triplet charged fermion mass $m_{\Sigma^\pm}$ in the Type-III Seesaw model~\cite{Foot:1988aq} to be $> 840$~GeV.}}

\section{The MoEDAL detector}
\label{sec:moedal}

The MoEDAL detector~\cite{Acharya:2014nyr} is located at Point 8 of the LHC, around the interaction point in the VErtex LOcator
(VELO) cavern of the LHCb detector. MoEDAL is largely a passive LHC detector, and the subdetector
system of principal relevance for this analysis is comprised of a large array ({$120$~m$^2$}) of 
Nuclear Track Detector (NTD) stacks composed of CR39 and Makrofol plastic surrounding the 
{interaction} region. MoEDAL also has paramagnetic trapping volumes (MMTs) that can capture 
highly-ionizing electrically- and magnetically-charged particles. The possible decays of trapped long-lived 
electrically-charged particles can be monitored at a remote facility, and magnetically-charged particles
are monitored at the ETH Zurich SQUID Magnetometer Facility. 
MoEDAL also incorporates an array of TimePix pixel devices that serves as a real-time 
system for monitoring highly-ionizing backgrounds in the cavern. We do not consider the MMTs
and TimePix devices in this analysis, but focus on the capabilities of the NTD stacks.

The NTD elements of the MoEDAL detector are passive, not needing a trigger, electronic readout, 
high-voltage or gas system. They are calibrated directly for highly-ionizing particles
by using heavy-ion beams. Thus, they complement the main LHC detectors, ATLAS and CMS,
which are not optimized for detecting heavily-ionizing {particles.
The NTD array has a low threshold and can detect particles with an ionization
level $Z/\beta \sim 7$}, where $Z$ is the electric charge and $\beta$ is the velocity of the particle. 
The charge resolution of the plastic NTDs is better than $0.05 e$, where $e$ is the electric charge.
The TDR NTD array is supplemented by a Very High Charge Catcher (VHCC) subdetector with threshold 
$Z/\beta \sim 50$ applied directly to the outside of the VELO detector housing so as to 
increase the geometrical acceptance for magnetic monopoles, which does not play a role in our analysis.

{After the NTD sheets are etched they must be scanned using optical microscopes, 
since the feature (etch-pit) sizes lie in the range 20 to 50~$\mu$m. A computer-controlled 
optical scanning microscope system will be deployed for MoEDAL data taking 
during Run 3. The system will be controlled by dedicated artificial intelligence (AI) 
software that is designed to recognize signal etch-pits in the presence of 
beam-induced backgrounds.}

The multi-sheet structure of the 
NTD stack enables the measurements of individual pits to be combined to define a precise trajectory 
and effective $Z/\beta$ values that measure the change in ionization energy loss 
as the particle loses energy during its passage through the NTD detector stack, 
demonstrating that the track comes from the interaction region and has $dE/dx$ values 
consistent with a heavily-ionizing electrically-charged particle.

These characteristics of the NTD system ensure that such a particle can be
detected with high efficiency and confidence within the geometric acceptance of the detector.

\section{Analysis framework}
\label{sec:framework}

In this study we consider the pair production of singly- and doubly-charged particles at the LHC,
$p p \to Y \overline Y + X$, where $\overline Y$ represents the antiparticle of $Y$ (which is
same as $Y$ if it is self-conjugate), and
$X$ represents soft particles originating from the beam remnants and
initial- and final-state QCD radiation.
The particle $Y$ may be a supersymmetric particle, specifically
an R-hadron containing
the strongly-interacting
$\tilde g, \tilde q, \tilde t$, a chargino $\tilde \chi^\pm$ or a charged slepton $\tilde \ell^\pm $,
or a doubly-charged particle with spin-0, $S^{++}$, or spin-1/2, $f^{++}$.

As already discussed, the MoEDAL detector is insensitive to electrically-neutral particle{s}
and a coloured supersymmetric particle must be hadronized into a charged R-hadron to be detected.
Since this probability $\kappa$ (the charged vs neutral R-hadron fraction) is not well understood \cite{ATLAS:2019duq},
we vary this parameter over the range $\kappa \in [0.5, 0.7]$. 
The distances between the interaction point and the MoEDALS's NTD panels are on average $\sim 2$\,m,
so MoEDAL is sensitive to particles $Y$ with lifetimes $\tau$ longer than ${\cal O}(1)$\,m/$c$,
otherwise the detection probability is exponentially suppressed.
In this study we treat $\tau$ as a free parameter and in the next Section we identify the region
of the mass vs lifetime plane for each choice of $Y$ that can be probed by MoEDAL with Run~3 data.   

As also discussed in the previous Section, the MoEDAL detector is essentially {free of SM backgrounds},
{and} we are {therefore} interested in the regions of parameter space where the numbers of expected signal events 
detected by MoEDAL are $N_{\rm sig} \geq 1$ and 2, which we {define} as the 
thresholds for ``evidence'' and ``discovery''.
The numbers of expected signal events are estimated by Monte Carlo simulation 
using the following formula:
\begin{equation}
\hspace{-3mm}
N_{\rm sig}(m,\tau) = \sigma(m) \cdot L \cdot \left\langle \sum_i \Theta(\beta_{\rm th} - \beta_i) P(\vec{\beta}_i,\tau) \right\rangle_{\rm MC}\,,
\label{eq:nsig}
\end{equation}
where $\sigma(m)$ is the production cross-section
as a function of the target particle's mass, $m$,
$L$ is the integrated luminosity,  $\vec{\beta}$ is the particle's three-velocity,
and $\langle \cdots \rangle_{\rm MC}$ represents the Monte Carlo average.
The summation over $i$ includes the two target particles $Y$ and $\overline Y$ in the event.
The Heaviside step function ($\Theta(x) = 1$ for $x > 0$ and 0 otherwise)
models the fact that the NTDs are capable to detect a particle with ionization level
higher than some threshold, $Z/\beta > 1/\beta_{\rm th}$.

The threshold velocity $\beta_{\rm th}$ for detection depends in principle on
the incident angle between the direction of the particle and the NTD panel.
This effect was significant in Run~2 of the LHC, because the NTD panels were not ideally oriented
(see~\cite{Felea:2020cvf} for a study of this effect).
However, in Run~3 all the NTD panels will be facing the interaction point and the incident angles
are always close to optimal.
We therefore use a constant threshold velocity $\beta_{\rm th}$ in this study for Run~3.
We take $\beta_{\rm th} = 0.15$ (0.3) for singly- (doubly-)charged particles.

We denote the probability for the target particle with three-velocity $\vec{\beta}$ 
and lifetime $\tau$ to survive 
and hit an NTD panel by $P(\vec{\beta}_i, \tau)$, which can be expressed as
\begin{equation}
    P(\vec{\beta}, \tau) = \epsilon(\vec{\beta}) \cdot \exp\left(-\frac{l(\vec{\beta})}{\gamma \beta c \tau} \right)\,. 
\end{equation}
In this expression $\epsilon(\vec{\beta}) = 1$ if there is an NTD panel in the direction of $\vec{\beta}$
and 0 otherwise.
The exponential factor represents the probability that the target particle does not decay before
reaching an NTD panel, where
$l(\vec{\beta})$ is the distance between the interaction point and
the NTD panel in the direction of $\vec{\beta}$,
$\tau$ is the lifetime and $\gamma = (1 - \beta^2)^{-1/2}$.  

In order to evaluate the Monte Carlo average in Eq.~\eqref{eq:nsig}
we used {\tt Madgraph5\_aMC\,\,v-2.6.7} \cite{MG}, and
implemented several types of doubly-charged particles
using {\tt FeynRules\,\,2} \cite{Christensen:2008py}.


\section{Singly-charged sparticles}
\label{sec:susy}

\subsection{Motivations for candidates}

Candidates for the long-lived sparticle include the gluino, squarks, 
Winos and Higgsinos, and sleptons. 

The gluino is not
detectable directly, since it hadronizes very soon after
production to form colour-neutral hadrons. The colour of the
gluino may be neutralized by a gluon, forming a neutral
hadron, or by a colour-octet quark pair. We focus on the
latter possibility, assuming that the gluino-gluon state is
heavier. We further assume that gluino/first-generation
quark states $\tilde g u \bar u$,$\tilde g d \bar d$, 
$\tilde g u \bar d$ and $\tilde g u \bar u$ are the lightest.
Depending on the details of superhadronic spectroscopy, the
lightest of all the gluino/quark states may be neutral or
charged. In the former case MoEDAL would not see a signal:
we consider here the latter case.

We consider separately the possibilities that the long-lived
sparticle is the lightest squark, which may be a partner of 
one of the 5 lightest quarks, or a stop. We assume that in the
former case the differences between these squark masses
would be relatively small, with the stops much heavier,
whereas in the light stop case the other squarks would be much heavier.
In both cases, we assume that the other squark species decay 
into the lightest one on a time scale {$\ll {\cal O}(1)$~m/$c$}. The
resultant long-lived squark would also appear as a bound state, of
which the lightest is expected to be that with a
first-generation antiquark, e.g., $\tilde u \bar u$, $\tilde d \bar d$, 
$\tilde u \bar d$ or $\tilde d \bar u$ if the lightest
squark is associated with the first generation. In the former 
two cases the lightest squark hadron would be neutral, and
MoEDAL would not see a signal: we assume here one of the 
latter cases of a charged long-lived particle.

Another possibility is that the long-lived sparticle is some
mixture of Wino, $\widetilde W$, and Higgsino, $\tilde h$, and we consider the cases
where one or the other component dominates production. Decays into
the lightest $\widetilde W/\tilde h$ combination would again occur on a time scale 
{$\ll {\cal O}(1)$~m/$c$}, and this combination could be either neutral
or charged: we consider the latter case here.

We consider finally the possibility that the long-lived sparticle
is a {charged} slepton, $\tilde \ell$, and our results apply to any flavour of slepton, assuming that the other slepton flavours
are much heavier. The lightest slepton might be the 
supersymmetric partner of either a left- or right-handed
lepton, and we consider both possibilities here, making the 
conservative assumption
that the heavier one is heavy enough to be effectively decoupled.
{In what follows we refer to the lightest slepton as the $\tilde \tau_{L/R}$, since
the tau slepton is the lightest in many supersymmetric models.
However, it should be noted that the numerical results for $\tau_{L/R}$ are also applicable to the other sleptons, namely
$\tilde e_{L/R}$ and $\tilde \mu_{L/R}$, as long as they are the lightest 
and other states are decoupled.
}
We recall that the detection with MoEDAL of long-lived sleptons via the cascade decays of long-lived gluinos was studied in~\cite{Felea:2020cvf, Sakurai:2019bac}, whereas in this Section we consider the direct pair production of long-lived sleptons.

\subsection{Run~3 projections for supersymmetric particles}
\label{sec:res_susy}

\begin{figure}[]
\centering
\includegraphics[width=0.425\textwidth]{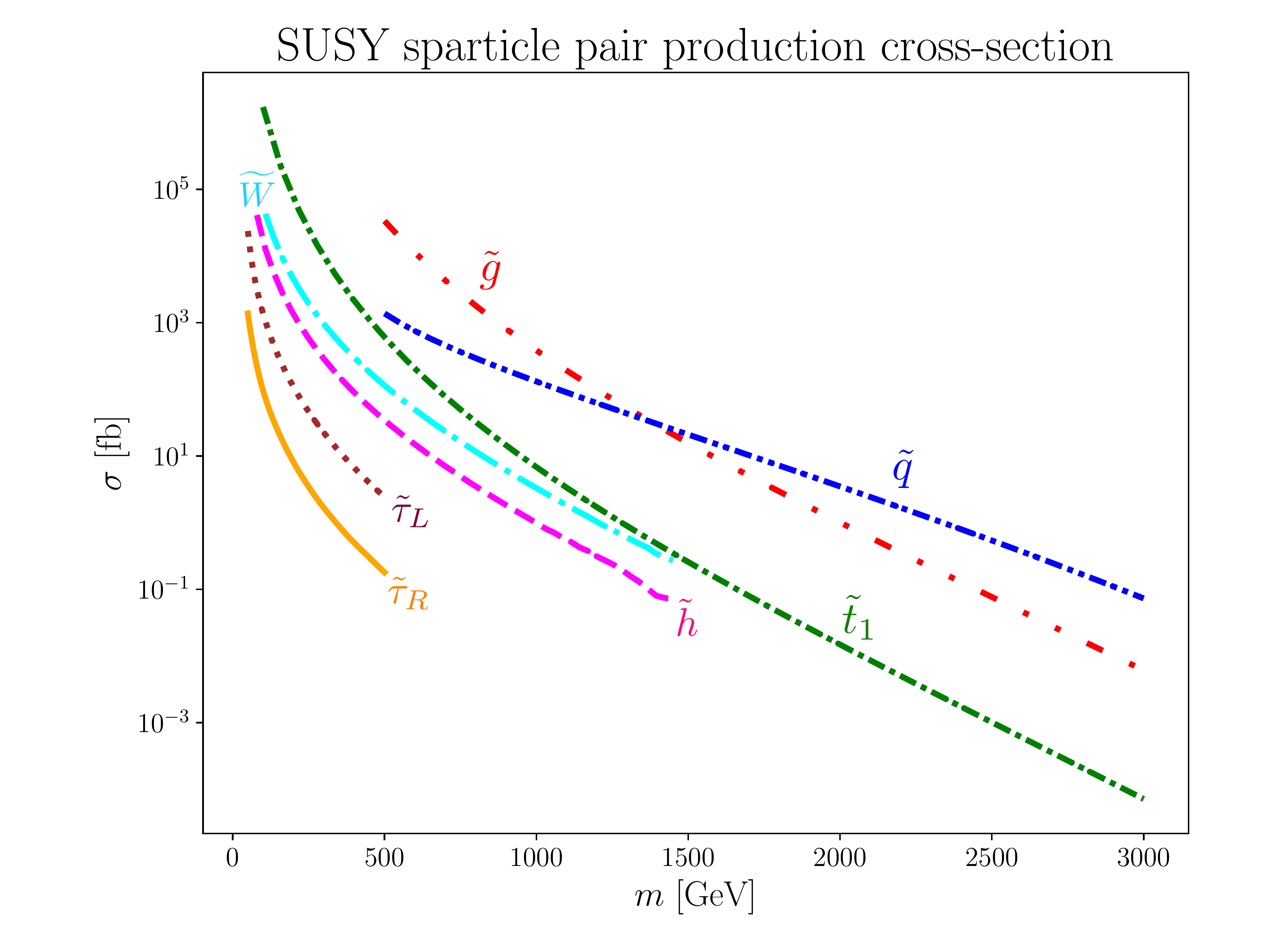}
\includegraphics[width=0.475\textwidth]{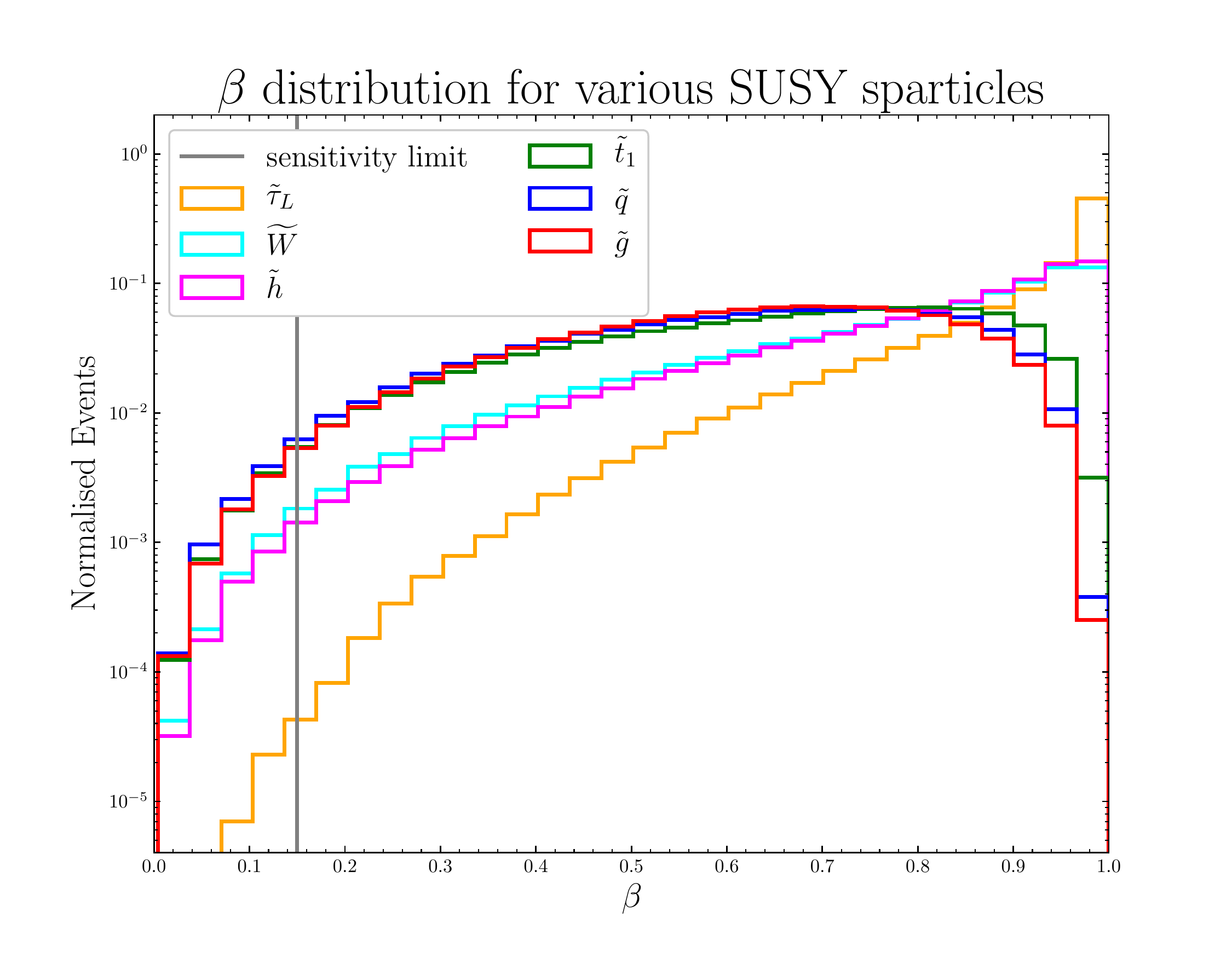}
\vspace{-1cm}
\caption{\it \small Upper panel: the production cross-sections for various sparticle species.
For the coloured particles  ($\tilde g$, $\tilde q$ and $\tilde t_1$),
we show NNLO$_{\rm Approx}$+NNLL cross-sections~\cite{susy-xs-wg}, 
while the cross-sections plotted for the weakly-interacting particles ($\widetilde W$, $\tilde h$ and $\tilde \tau_{L/R}$)
are calculated to NLO+NLL accuracy~\cite{nnllfast}.
Lower panel: the velocity distributions for various sparticle species.
The following representative masses were chosen:
$\tilde g$: 1010~GeV, $\tilde q$: 920~GeV, $\tilde t_1$: 720~GeV,
$\widetilde W$: 300~GeV, $\tilde \tau_L$: 80 GeV.}
\label{fig:susy}
\end{figure}

We begin the discussion of MoEDAL's projected sensitivity at Run~3
for singly-charged long-lived supersymmetric particles,
showing in the upper panel of Fig.~\ref{fig:susy}
the 13 TeV cross-sections for pair production of 
various supersymmetric particle species.
The cross-sections are all taken from the
{\tt LHC SUSY Cross Section Working Group}~\cite{susy-xs-wg},
except for $\tilde \tau_L$, for which we used {\tt Resummino\,\,2.0.1} \cite{resummino}. 
The cross-sections for coloured supersymmetric particles 
($\tilde g$, $\tilde q$ and $\tilde t_1$)
are computed including approximate next-to-next-to-leading order (NNLO$_{\rm Approx}$) supersymmetric QCD corrections and the 
resummation of soft gluon emission at next-to-next-to-leading-logarithmic
(NNLL) accuracy \cite{nnllfast},
whereas those for the weakly-interacting supersymmetric particles ($\widetilde W$, $\tilde h$ and $\tilde \tau_{L/R}$)
are calculated to NLO+NLL accuracy. 

The curve for $\tilde q$ {in} the upper panel of Fig.~\ref{fig:susy}
is the sum of the cross-sections for both the left- and right-handed versions of all 5 light-flavoured squarks,
calculated assuming $m_{\tilde g} = 3$ TeV.
In the cases of electroweakly-interacting sparticles,
the cross-sections are summed over all triplet (doublet) components for $\widetilde W$
($\tilde h$ and $\tilde \tau_L$).
Namely, we included $\widetilde W^\pm \widetilde W^0$ and $\widetilde W^+ \widetilde W^-$ production for Winos,
$\tilde h^+ \tilde h^-$, $\tilde h^0_{1,2} \tilde h^\pm$ and $\tilde h_1^0 \tilde h_2^0$ production for Higgsino
and 
$\tilde \tau_L^+ \tilde \tau_L^-$,
$\tilde \nu_\tau \tilde \tau_L^\pm$ and
$\tilde \nu_\tau \tilde \nu_\tau$ 
for producing the left-handed slepton,
assuming that the heavier components of the multiplets decay promptly into the lightest 
charged partner, which we assume to be long-lived.\footnote{
{Charged Winos are usually heavier than the neutral one by an amount $> m_{\pi^\pm}$,
due to radiative corrections. However, the mass ordering can be reversed by tuning the mixing in the chargino and neutralino sectors, which is the scenario considered in this study.
} 
}

As expected, one can see in Fig.~\ref{fig:susy} that the coloured sparticles have the largest cross-sections,
while sleptons $\tilde \tau_{L/R}$ have smaller cross-sections.
For cross-sections to be above 1 pb, the masses have to be smaller than around
$1$ TeV ($\tilde g$), $700$ GeV ($\tilde q, \tilde t_1$), $400$ GeV ($\widetilde W$, $\tilde h$)
and $200$ GeV ($\tilde \tau_{L/R}$).

As already mentioned, the production velocity of a
singly-charged particle must be $< \beta_{\rm th} = 0.15$
for it to be detectable by MoEDAL's NTDs.
In the lower panel of Fig.~\ref{fig:susy} we
show the normalised velocity distributions
for various sparticle species, as computed with {\tt MadGraph5\_{aMC}}. 
We took the following representative masses for this purpose:
$\tilde g$: 1010~GeV, $\tilde q$: 920~GeV, $\tilde t_1$: 720~GeV,
$\widetilde W$: 300~GeV, $\tilde \tau_L$: 80~GeV.
The velocity distribution of the
right-handed slepton $\tilde \tau_R$ (not shown)
is similar to that shown for the $\tilde \tau_L$.

We see that the coloured sparticles have,
in general, much lower velocities  than the
weakly-interacting sparticles.
One reason for this is that the masses used in this calculation are larger for coloured particles
than for weakly-interacting ones, so their production is more central.
Another effect is that the production of coloured particles may be 
dominated by the gluon-gluon-initiated $t$-channel process.
On the other hand, the production of weakly-interacting particles is
largely dominated by the Drell-Yan $s$-channel process from a quark-antiquark 
initial state.
We also observe that the fermionic particles $\widetilde W$ and $\tilde h$,
have much lower velocities than the scalar particle $\tilde \tau_{L}$ ($\tilde \tau_{R}$)
on average.
This is because the $s$-channel process is mediated by spin-1 gauge bosons ($\gamma/Z/W^\pm$),
and the pair-production rate for scalar particles vanishes in the limit of $\beta \to 0$
because of $p$-wave suppression, which is absent for 
the pair production of the fermionic particles $\widetilde W$ and $\tilde h$.

The grey vertical line in the lower panel of Fig.~\ref{fig:susy}
marks the detection threshold velocity $\beta_{\rm th} = 0.15$,
and the portions of the distributions to the left of this line can be detected.
We see that the detection efficiency is highest for coloured supersymmetric particles,
much lower for the slepton,
and intermediate for $\widetilde W$ and $\tilde h$.

\begin{figure}[]
\centering
\includegraphics[width=0.45\textwidth]{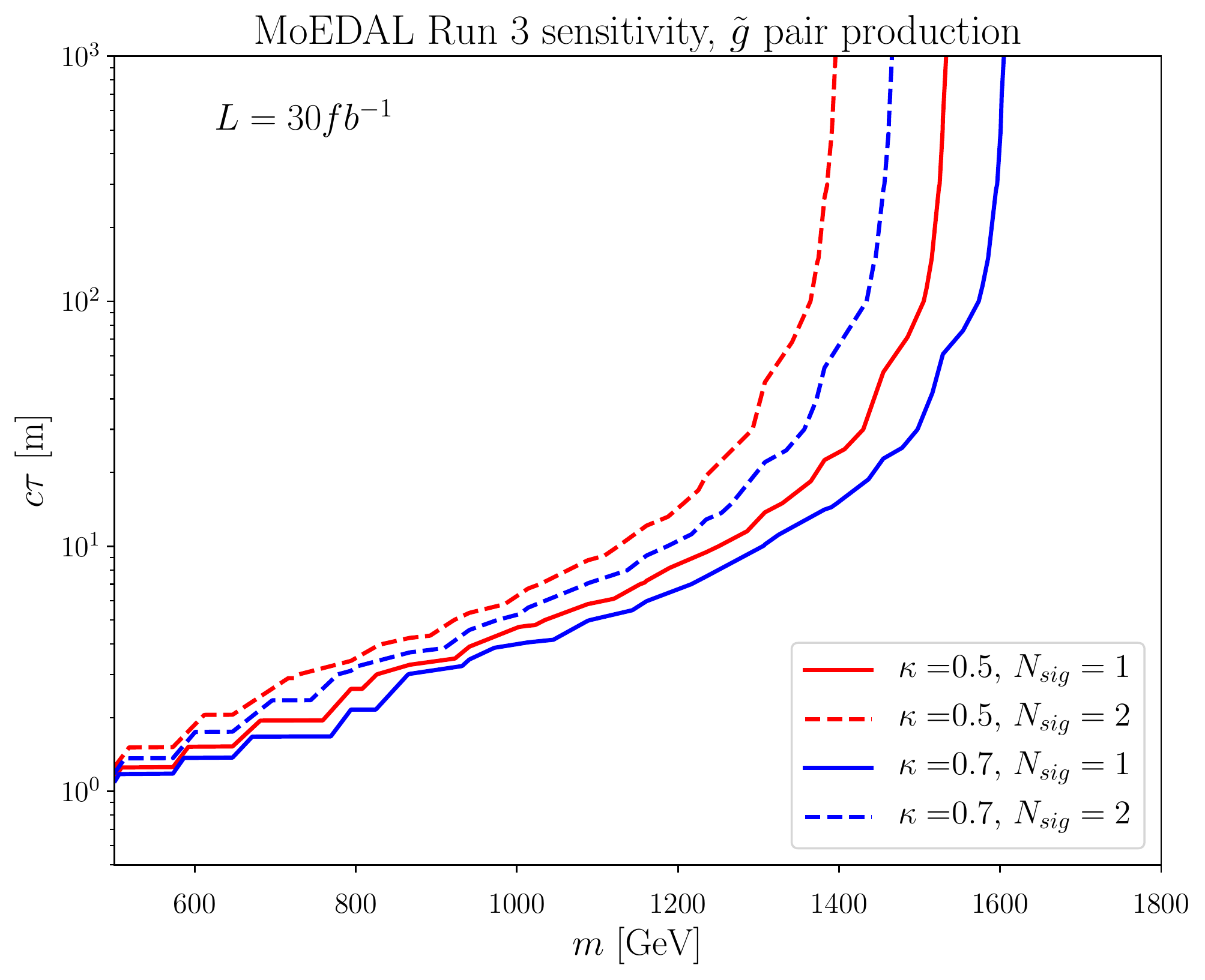}
\includegraphics[width=0.45\textwidth]{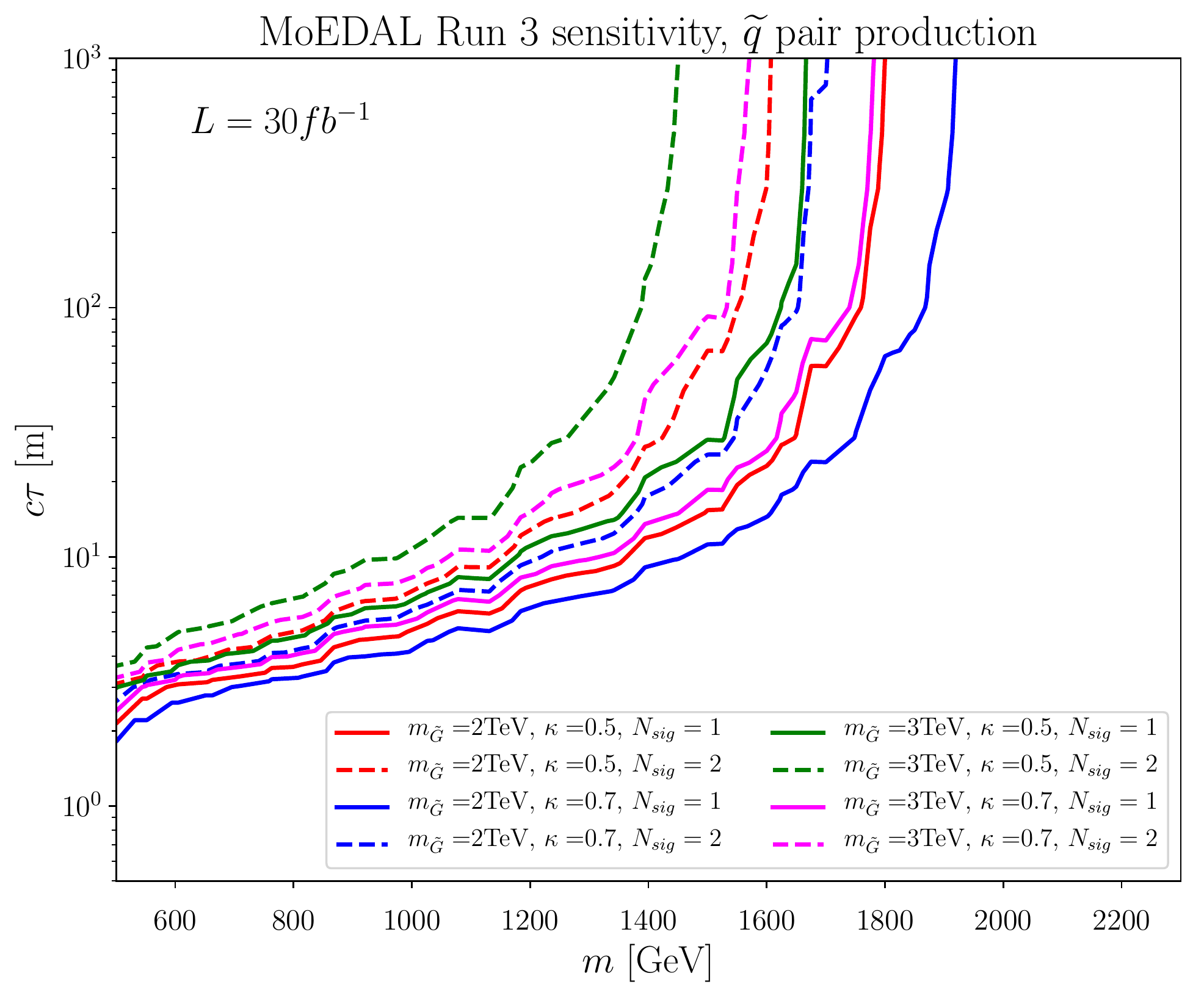}
\includegraphics[width=0.45\textwidth]{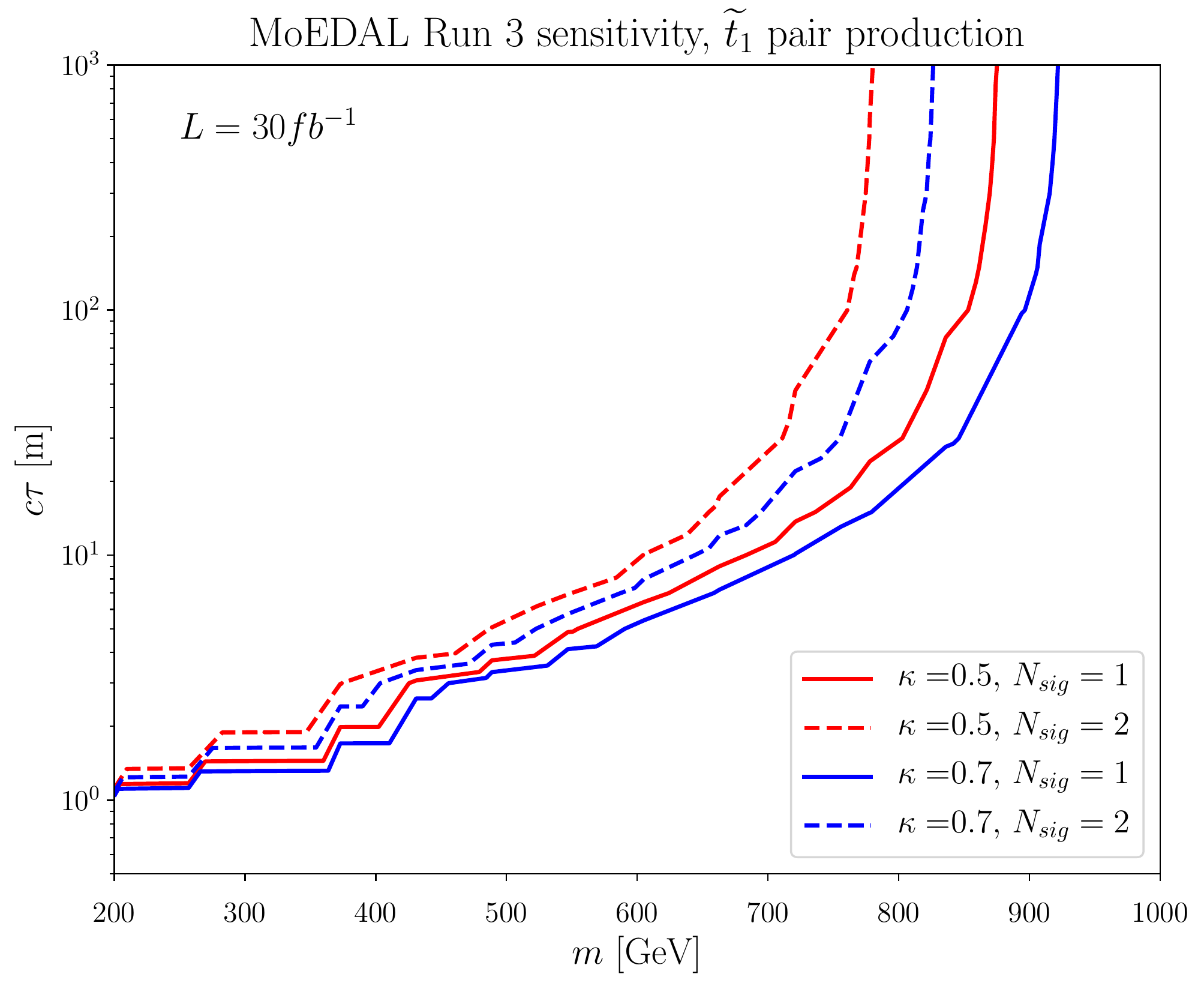}
\vspace{-0.5cm}
\caption{\it \small The expected sensitivities of MoEDAL for various long-lived strongly-interacting sparticle species, assuming 30 fb$^{-1}$ of integrated luminosity.
Top panel: the gluino, middle panel: squarks, bottom panel: the stop squark.}
\label{fig:strong}
\end{figure}

We now discuss the expected sensitivities of MoEDAL to long-lived supersymmetric particles 
in Run~3 of the LHC, assuming an integrated luminosity of 30 fb$^{-1}$.
In Fig.~\ref{fig:strong} we show the contours of $N_{\rm sig} = 1$ (solid) and $2$ (dashed),
corresponding to the thresholds for ``evidence" and ``discovery", respectively,
for strongly-interacting sparticles in mass versus $c\tau$ planes.
The top panel shows the sensitivities for gluinos with
the red and blue contours corresponding to the charged R-hadron fractions of
$\kappa = 0.5$ and 0.7, respectively.
Under the optimistic assumptions $\kappa = 0.7$ and $c \tau \gtrsim 100$m, 
MoEDAL is expected to see 1 (2) signal event(s) for 
{
$m_{\tilde g} \simeq 1600$ (1470) GeV,
}
while in the more conservative case $\kappa = 0.5$, 
the mass reach is 
{$\sim 1530$ (1400) GeV for $N_{\rm sig} = 1$ (2).}

The middle panel of Fig.~\ref{fig:strong} shows the MoEDAL sensitivity to light-flavour squarks.
Since the production cross-section depends on $m_{\tilde g}$,
we show results for two gluino masses, $m_{\tilde g} = 2$~TeV for $\kappa = 0.5$ (red) and $\kappa = 0.7$
(blue) and 3~TeV for $\kappa = 0.5$ (green) and $\kappa = 0.7$
(magenta).
We see from the plot that the mass reach is greater for the smaller gluino mass, 
since the cross-section for light-flavour squark production is larger in this case.
In the most optimistic case ($m_{\tilde g} = 2$~TeV, $\kappa = 0.7$, $c\tau \gtrsim 100$m),
MoEDAL could detect the gluino up to
{
$\sim 1920$ (1700) GeV for $N_{\rm sig} = 1$ (2).
}
On the other hand for $m_{\tilde g} = 3$ TeV and $\kappa = 0.5$, the mass reach 
{
is $\sim 1670$ (1450) GeV for $N_{\rm sig} = 1$ (2).}

The Run~3 sensitivity for the lighter stop, $\tilde t_1$, is presented in the bottom panel
of Fig.~\ref{fig:strong}, where
the convention for the line-styles is the same as in the top panel.
We see that a long-lived $\tilde t_1$ could be probed by MoEDAL up to 
{
$m_{\tilde t_1} \sim 920$ ($N_{\rm sig} = 1$) and 830 ($N_{\rm sig} = 2$) GeV
for $\kappa = 0.7$,
while the reach is $\sim 870$ ($N_{\rm sig} = 1$) and 780 ($N_{\rm sig} = 2$) GeV
for $\kappa = 0.5$.
}

\begin{figure}[]
\centering
\includegraphics[width=0.45\textwidth]{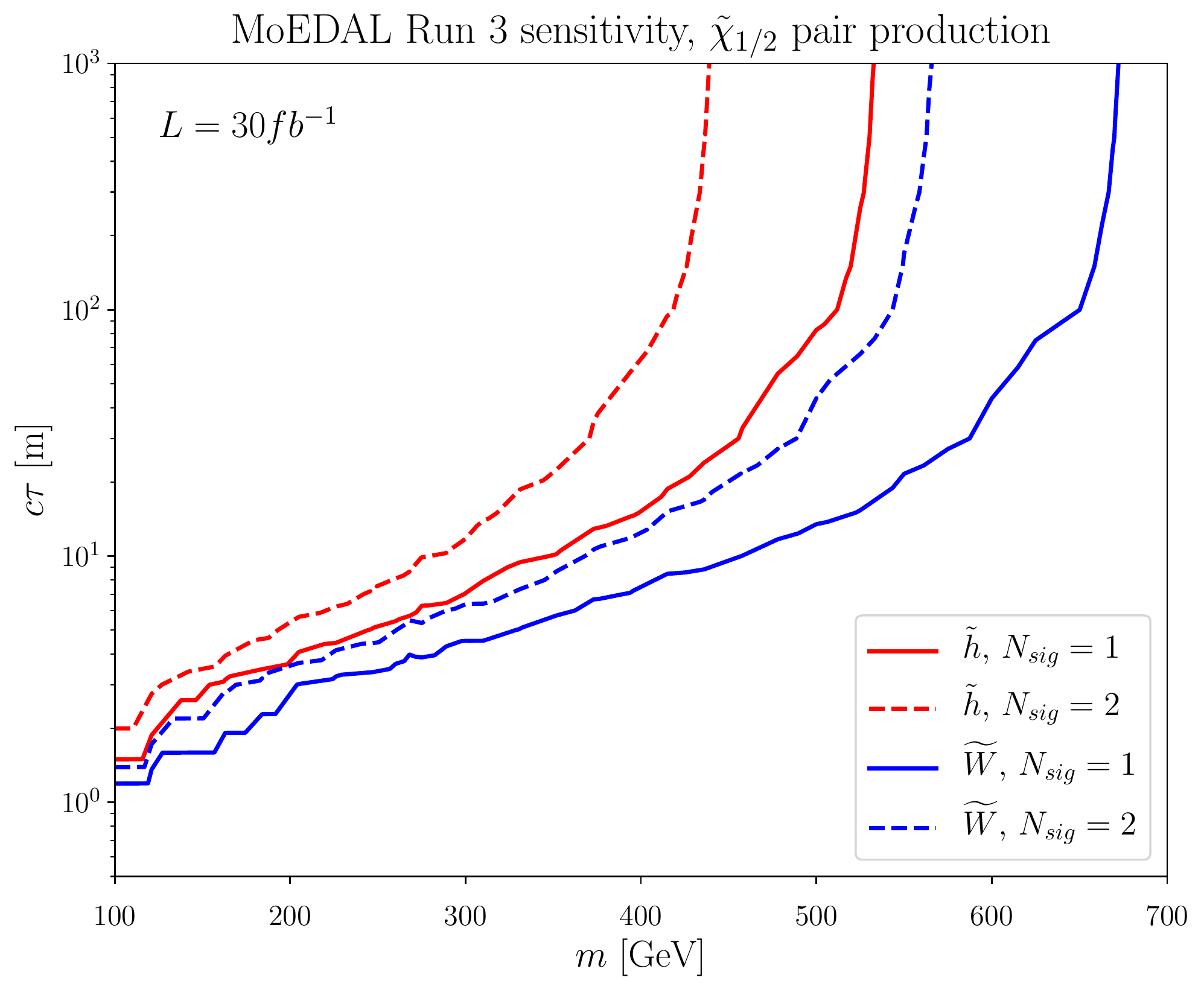}
\includegraphics[width=0.45\textwidth]{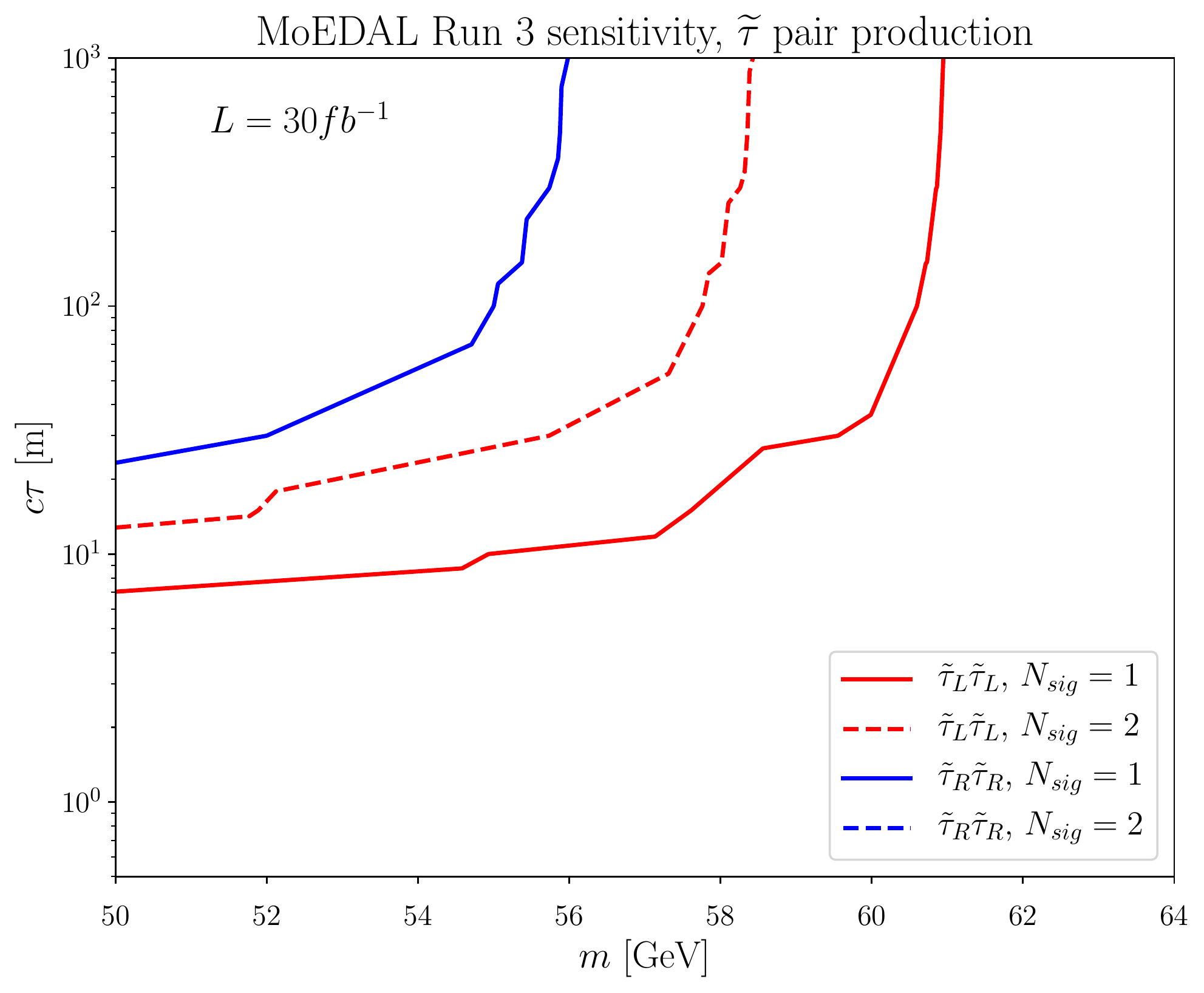}
\vspace{-0.5cm}
\caption{\it \small The expected sensitivities of MoEDAL for various long-lived electroweakly-interacting sparticle species, assuming 30 fb$^{-1}$ of integrated luminosity.
Upper panel: charginos, lower panel: sleptons.}
\label{fig:weak}
\end{figure}

We now turn to discuss the expected sensitivities for weakly-interacting sparticle species
shown in Fig.~\ref{fig:weak}. The Run~3 projections for MoEDAL searches
for a long-lived Wino (blue) and Higgsino (red) are shown in the top panel.
We see that MoEDAL could probe the Wino up to
{
$m_{\widetilde W} \sim 670$ (570) GeV 
}
for $N_{\rm sig} = 1$ (2) if $c\tau \gtrsim 100$\,m,
whereas the mass reach for the Higgsino with $c\tau \gtrsim 100$\,m is 
{
$m_{\tilde h} \sim$ 530 (430) GeV
}
for $N_{\rm sig} = 1$ (2).
The higher mass reach for the Wino is obtained because the cross-section is larger 
thanks to {its} larger SU(2) charge.

The bottom panel of Fig.~\ref{fig:weak} shows the MoEDAL Run~3 sensitivities for
long-lived sleptons, $\tilde \tau_R$ (blue) and $\tilde \tau_L$ (red).
As can be seen, the mass reach is very low:
{
it is 61 (58) GeV
}
for the meta-stable $\tilde \tau_L$ with $N_{\rm sig} = 1$ (2),
and
{56 GeV} for $\tilde \tau_R$ with $N_{\rm sig} {=} 1$
(there is no corresponding contour for $N_{\rm sig} {=} 2$ for $\tilde \tau_R$.).
All of these reaches are {below limits established at LEP~\cite{PDG}} and already excluded.
The reason for the low sensitivity is twofold.
First, as discussed above the cross-sections for slepton production are very low, 
since production is mediated by spin-1 gauge bosons in the
$s$-channel process and the production rate is velocity suppressed.
Secondly, for the same reason the produced sleptons must have larger velocities
than the other sparticle species examined here, as was also discussed previously.
We conclude that, since MoEDAL targets highly-ionizing particles with $Z/\beta \gtrsim 7$,
it is not sensitive to sleptons.

\subsection{Comparison with the existing searches}
\label{sec:comp_susy}

We now compare the prospective MoEDAL sensitivities at Run~3 with the existing mass limits
for metastable supersymmetric particles.
Several different types of long-lived signatures have been discussed 
in the literature and searched for by ATLAS and CMS.
For example, if gluinos have a lifetime of $1-100$\,mm$/c$ and
decay into dijets, they can be searched for by reconstructing 
the displaced vertices originating from meta-stable
gluino decay~\cite{Aaboud:2017iio, Sirunyan:2018pwn}.
Another example is a disappearing track signature from long-lived winos
with $\tau \sim 1 - 10$\,cm$/c$.
The long-lived signature emerges when a charged wino is produced and 
decays into a nearly mass-degenerate neutral partner inside one of
the silicon track detectors~\cite{Aaboud:2017mpt, Sirunyan:2020pjd}.
Although the aforementioned searches are powerful for
specific scenarios, they are not relevant 
for particles with longer lifetimes,
or when the dominant decay modes differ from those assumed in the analyses.
Since the MoEDAL search is independent of the long-lived particle decay mode
after passing through the NTDs, it avoids the model dependence of the above constraints.

There is, however, one type of search whose constraints are less model-dependent.
These are heavy stable charged particle (HSCP) searches,
which rely on large ionization energy loss $dE/dx$, {the MoEDAL signature considered here},
and time-of-flight (ToF) measurements,
both of which are independent of the nature of decays.~\footnote{A
scenario that can relax the constraint from  
HSCP analyses has been proposed and studied in \cite{Aaboud:2019trc}.}
We therefore focus now on the constraints obtained from the HSCP searches.

\begin{table}[t!]
  \begin{center}
    \renewcommand{\arraystretch}{1.3}
  \begin{tabular}{|c||c|c|c|} 
\hline
 & MoEDAL   & (ATLAS)& (CMS)  \\ 
\hline
$\tilde g$ & {1600} & {(2000)} & 
{(1500)}  \\ 
\hline
$\tilde q$ & {1920} & {((2310))} & -  \\ 
\hline
$\tilde t$ &  {920} &  {(1350)} & 
{(1000)}\\ 
\hline 
$\widetilde W$ &  {670} &  {(1090)} & -  \\ 
\hline 
$\widetilde h$ & {530} & {((1170))} & -  \\ 
\hline 
$\widetilde \tau$ & {61} & {(430)} & 
{(230)}\\
\hline
  \end{tabular}
\end{center}
\vspace{-2mm}
\caption{\it \small Comparison between the MoEDAL mass reaches 
at Run~3 for $N_{\rm sig} = 1$, $\kappa=0.7$ 
and (in parentheses) the current (95\,\% CL) mass bounds on 
several long-lived supersymmetric particle candidates 
obtained (estimated in the cases of $\tilde q$ and $\tilde h$,
in double parentheses) 
from the ATLAS heavy stable charged particle search 
with $L = 36.1$ fb$^{-1}$ \cite{Aaboud:2019trc} (second column) and CMS~\cite{Khachatryan:2016sfv} (third column) under the model-dependent
assumptions described there and in the text.
All masses are in GeV.
}
  \label{tab:susy_limit}
\end{table}

In Table~\ref{tab:susy_limit} we summarise
the prospective MoEDAL mass reaches at Run~3 for $N_{\rm sig} = 1$, $\kappa=0.7$
and compare them with the current (95\% CL) mass limits 
obtained by the most recent ATLAS HSCP analysis with $L = 36.1$ fb$^{-1}$~\cite{Aaboud:2019trc}.
The ATLAS Collaboration interpreted their results for
long-lived gluino, stop, Wino {and stau} candidates,
and derived the lower bounds on their masses
shown in parentheses in the third column of Table~\ref{tab:susy_limit}.
At the time of writing, the most recent CMS HSCP analysis~\cite{Khachatryan:2016sfv} 
is based on a smaller set of data 
($L = 2.5$ fb$^{-1}$) and its limits on the gluino, stop and {stau} are weaker than those of ATLAS, 
as seen in the last column of Table~\ref{tab:susy_limit}.

The ATLAS constraints on the $\tilde g$, $\tilde t$ and $\widetilde W$
long-lived particle candidates are nominally stronger than the prospective
reaches of MoEDAL with Run~3 data. However, the ATLAS
selection cuts for this search may lead to gaps in the full phase space coverage  that
can be avoided by MoEDAL.
As summarised in Table~1 of~\cite{Aaboud:2019trc}, ATLAS defined
5 signal regions (SRs), all of which incorporated $E_{\rm T}^{\rm miss}$
triggers (varying between 70 and 110~GeV) or single (isolated) high-momentum muons.~\footnote{Similarly,
the CMS analysis~\cite{Khachatryan:2016sfv} incorporates an $E_{\rm T}^{\rm miss} > 170$~GeV requirement.}
Depending on the scenario for long-lived sparticles that is considered,
either or both of these auxiliary signatures may be absent.
For example, there would be no $E_{\rm T}^{\rm miss}$ signature in a scenario
with weak R-parity violation, so these events would need
to get selected by a muon-like signature in the detector.

We note also that in~\cite{Aaboud:2019trc} ATLAS did not interpret their result 
in scenarios with long-lived light-flavour squarks or Higgsinos.
In order to estimate the possible ATLAS mass reaches for these particles,
we recast the cross-section upper limits for sbottoms (Winos)
derived by ATLAS \cite{Aaboud:2019trc}.
Our recasting assumes that the detection efficiencies of the HSCP analysis 
would be similar for light-flavour squarks and bottom
squarks, and for Winos and Higgsinos.
Based on this assumption we derive the mass bounds 
{2310 (1170)}
GeV for squarks 
(Higgsinos) shown in double parentheses in Table~\ref{tab:susy_limit}.

Comparing the prospective sensitivities of MoEDAL for long-lived sparticles
with those of ATLAS and CMS, we note the following two points.
One is that the luminosity to be accumulated by MoEDAL is only about one fifth 
of that already accumulated by ATLAS or CMS.
This is because MoEDAL is located at Point 8 of the LHC
together with {the LHCb detector, which requires a restricted instantaneous luminosity so as to suppress 
pile-up}.~\footnote{{On the positive side, this restriction results in lower beam-related background in the MoEDAL NTD detectors.}}
Secondly, MoEDAL is sensitive only to particles with the ionization levels 
higher than $Z/\beta \gtrsim 7${, which reduces the acceptance for singly-charged HSCPs, in particular.}

{Finally, we comment on the case of an $SU(2)_L$-triplet fermion, $(\Sigma^\pm, \Sigma^0)$, in a Type-III seesaw model, as was briefly mentioned in Section \ref{sec:intro}.
In general, the mass degeneracy between
$\Sigma^\pm$ and $\Sigma^0$ 
is resolved due to radiative corrections
in such a way $m_{\Sigma^\pm} - m_{\Sigma^0} > m_\pi^\pm$,
which makes the lifetime of $\Sigma^\pm$ too short for an HSCP signal in MoEDAL.
However, if there are additional contributions to the mass splitting so that 
$m_{\Sigma^\pm} - m_{\Sigma^0} < m_\pi^\pm$,
the lifetime of $\Sigma^\pm$ would be long enough 
for detection of $\Sigma^\pm$ at MoEDAL to be possible. 
For $m_{\Sigma^0} > m_{\Sigma^\pm} + m_\pi^\pm$,
$\Sigma^0$ decays promptly to $\Sigma^\pm$ and 
the mass reach would be $\sim 670$ GeV and
the same as that for the Winos studied in this Section.
For $m_{\Sigma^0} + m_\pi^\pm > m_{\Sigma^\pm} > m_{\Sigma^0}$,
$\Sigma^\pm$
undergoes three-body decay to $\Sigma^0$ with a lifetime ${\cal O}(10^{-6})$\,s.
In this case, the $\Sigma^0$ does not contribute to the signal and
the signal yield would be reduced by roughly 1/2, since the dominant production mode is
$pp \to \Sigma^\pm \Sigma^0$.
We expect that the mass reach in this case would be similar to that for the Higgsino studied in this Section.
}

\section{Doubly-charged particles}

\label{sec:dcp}

\subsection{Motivations for candidates}
\label{sec:dcp_moti}

Doubly-charged scalars can arise in variety of scenarios for physics
beyond the standard model (BSM). 
{As mentioned in Section \ref{sec:intro}, the Type-II seesaw model can provide doubly-charged scalars $(H^{\pm\pm})$\cite{Schechter:1980gr, Magg:1980ut, Cheng:1980qt,Lazarides:1980nt, Mohapatra:1980yp, Lindner:2016bgg} that will be relevant for this analysis.}
  In addition to the Type-II seesaw model, several other BSM scenarios, namely the Left-Right model~\cite{Pati:1974yy, Mohapatra:1974hk,Senjanovic:1975rk}, the Georgi-Machacek (GM) model \cite{Georgi:1985nv,Chanowitz:1985ug,Gunion:1989ci,Gunion:1990dt,Ismail:2020zoz}, the 3-3-1 model~\cite{CiezaMontalvo:2006zt,Alves:2011kc} and the little Higgs model \cite{ArkaniHamed:2002qx} also predict doubly-charged scalars. The supersymmetric versions of these models can lead to 
  doubly-charged spin-1/2 Higgsinos. The minimal left-right supersymmetric model which is based on $SU(3)_C\times SU(2)L\times SU(2)_R \times U(1)_{B-L}$ is one such model\cite{Kuchimanchi:1993jg,Babu:2008ep,Francis:1990pi,Huitu:1993uv,Frank:2014kma}.
 {In addition to these particular well-studied models for multiply-charged scalars and fermions, such
 particles can also arise in the simplified models discussed in~\cite{Delgado:2011iz,Alloul:2013raa}.}

  In the pure Type-II seesaw model, the doubly-charged scalars $(H^{\pm\pm})$ belonging to the $SU(2)_L$-triplet scalar multiplet $\Delta $ couple to leptons and the $W$ boson,
  and the corresponding interaction strengths are controlled by the vacuum expectation value $(v_T)$ of the neutral component of the $\Delta$. This vev is related
  to the Majorana masses of the neutrinos: $M_\nu = \sqrt{2} Y_\nu v_T$, where $Y_\nu$ denotes the neutrino Yukawa couplings, and the collider limits on the
  the mass of the doubly{-}charged scalars depend on the value of $v_T$. For small $v_T \leq 10^{-4}$ GeV that corresponds to large $Y_\nu$, and assuming degenerate heavy scalar $(A,H,H^\pm, H^{\pm\pm})$ masses, 
  ${\rm Br}(H^{\pm\pm} \to \ell^\pm \ell^\pm) \approx 100 \% (\ell = e,\mu ) $. 
  Using the like-sign dilepton (LSD) final state, the current lower bound on  
  $m_{H^{\pm\pm}}$ varies between 770 GeV and 870 GeV \cite{Aaboud:2017qph} 
  at the $95\%$ CL from the direct search for a doubly-charged Higgs boson
  in the 13 TeV LHC run. The corresponding lower limit on $m_{H^{\pm\pm}}$ 
  changes significantly for $v_T \geq 10^{-4}$ GeV (small $Y_\nu$), 
  when the LSD decay mode of $H^{\pm\pm}$ is highly suppressed while several 
  competing decay modes of $H^{\pm\pm}$ start opening up, such as (i) a
  pair of heavy bosons $W^\pm W^\pm$, (ii) $W^\pm H^\pm $ and (iii) $ H^\pm H^\pm$,
  if these are kinematically accessible. The subsequent
  decays of $W$ and $H^\pm $ into various leptons and jets give rise to
  rather complicated final states. Due to the cascade nature of the final state, 
  the collider bound on the doubly-charged scalar is rather weak in this case. 
  
  The ATLAS collaboration studied the pair production of doubly-charged scalars 
  that subsequently decay into 
  pairs of $W$ bosons (assuming ${\rm Br}(H^{\pm\pm} \to W^\pm W^\pm ) \approx 100 \%)$ 
  in the 13 TeV LHC run. Non-observation of any signal beyond the standard model 
  background sets a new limit on the doubly-charged scalar mass. 
  Using a data sample from an integrated luminosity of $36.1~{\rm fb}^{-1}$, 
  the $m_{H^{\pm\pm}}$ has been excluded between 200-220 GeV at $95\%$ CL\cite{Aaboud:2018qcu}.
  
  The left-right (LR) symmetric model predicts two types of doubly-charged 
  scalars $H^{\pm\pm}_L$ and $H^{\pm\pm}_R$, corresponding to $SU(2)_L$ and 
  $SU(2)_R$ triplet scalars $\Delta_L$ and $\Delta_R$,
  respectively. As these two scalars belong to different gauge group, 
  their couplings with fermions and gauge bosons are distinctly different. This is
  reflected in their production rates at the LHC:
  $\sigma (pp \to H^{++}_L H^{--}_L + \cdots) \approx 2.3 \times \sigma (pp \to H^{++}_R H^{--}_R) + \cdots$,
  due to the different coupling strength of $H^{\pm\pm}_{L,R}$ with 
  the $Z$ boson \cite{Huitu:1996su, Aaboud:2017qph}.
  The ATLAS Collaboration looked for doubly-charged scalars in the LSD invariant mass
  distributions for the $e^\pm e^\pm $, $\mu^\pm \mu^\pm $ and $e^\pm \mu^\pm$ 
  final states, and also in final
  states with three or four leptons (only electrons and muons) in the 13 TeV LHC run with an integrated luminosity of $36.1~{\rm fb}^{-1}$. No significant excess over the standard model prediction was observed. As a result, lower mass limits were obtained for the $m_{H^{\pm\pm}_L}$ and $m_{H^{\pm\pm}_R}$, assuming 
  ${\rm Br}(H^{\pm\pm}_{L,R} \to \ell^\pm \ell^\pm) =100 \% (\ell = e, \mu)$. 
  The limit for $m_{H^{\pm\pm}_L}$ is same as that in the Type-II seesaw model.
  However, for the $m_{H^{\pm\pm}_R}$
  the observed lower limit varies between 660 GeV and 760 GeV at the
  $95\%$ CL\cite{Aaboud:2017qph}.
  
  In the GM model, the current LHC limit on $m_{H^{\pm\pm}}$ varies between 
  200 and 220 GeV at $95\%$ CL, as obtained using an integrated luminosity 
  of $36.1~{\rm fb}^{-1}$ in the 13 TeV LHC run by the 
  ATLAS Collaboration~\cite{Aaboud:2018qcu}. 
  
  One should note that in all these 
  search analyses only prompt decays of $H^{\pm\pm}_{L,R}$ scalars 
  $(c\tau < 10 \mu {\rm m})$ were considered~\cite{Aaboud:2017qph}. 
  Hence they are complementary to the long-lived particle search
  that is possible with MoEDAL. In 
  the Type-II seesaw model, there are certain regions of the
  $(v_T, m_{H^{\pm\pm}})$ parameter plane where the life-time of the 
  doubly-charged scalar can be much longer. If the $H^{\pm\pm}$ were to decay
  outside an LHC detector, 
  it would leave a heavily-ionizing charged track signal.
  Both the ATLAS and CMS collaborations have studied these signatures of such 
  heavy long-lived particles, as we discuss later.

\subsection{Run~3 projections for doubly-charged particles}
\label{sec:res_dbl}

We study here the prospective MoEDAL sensitivities to four types of doubly-charged particles:
(singlet, triplet) $\times$ (scalar, fermion).
For the singlet and triplet types, the weak gauge quantum number assignments are
$(SU(2)_L, U(1)_Y) = ({\bf 1}, 2)$ and $({\bf 3}, 1)$, respectively, and
all these particles are assumed to be colour singlets.

\begin{figure}[]
\centering
\vspace{-1cm}
\includegraphics[width=0.475\textwidth]{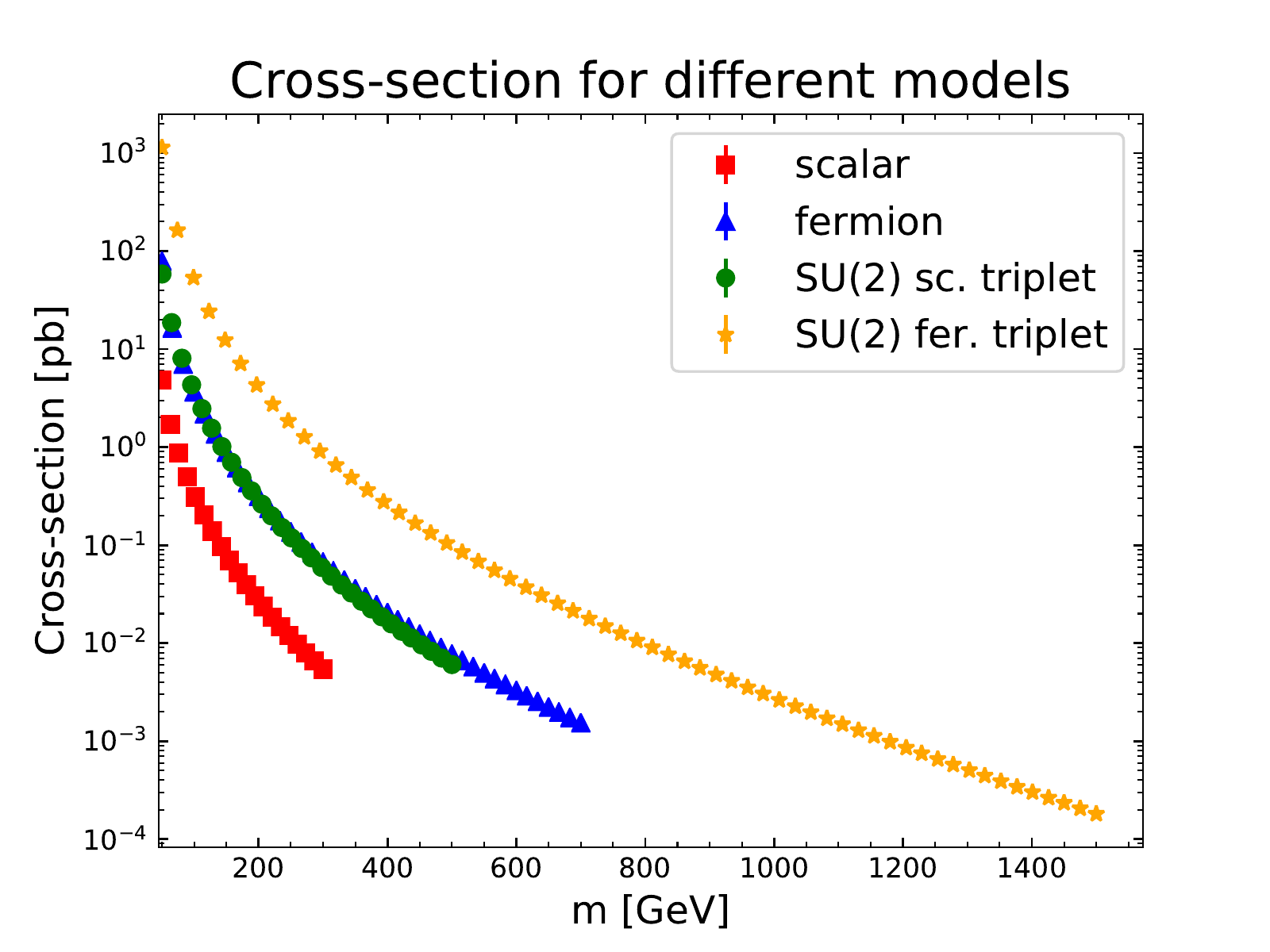}
\includegraphics[width=0.475\textwidth]{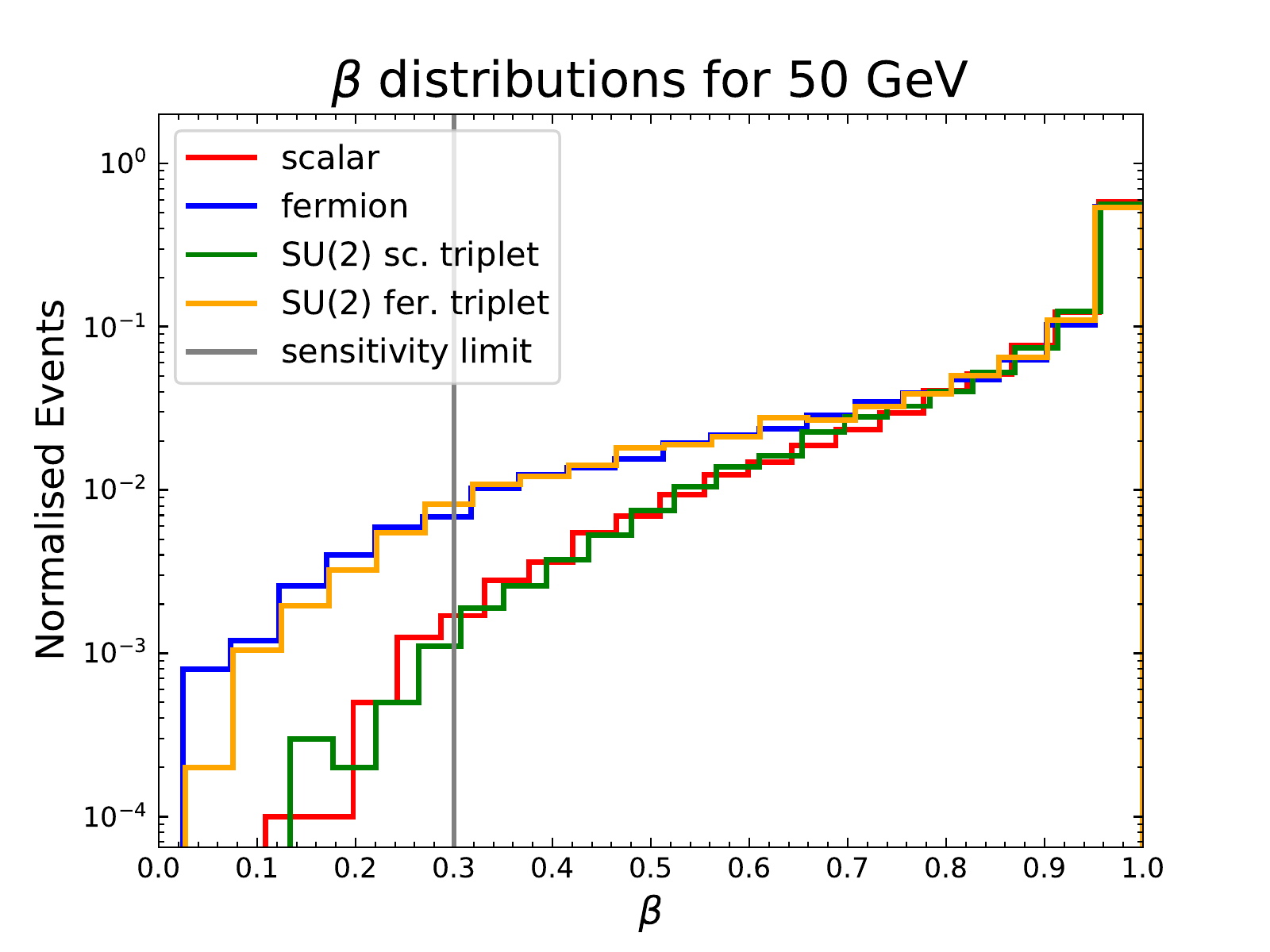}
\includegraphics[width=0.475\textwidth]{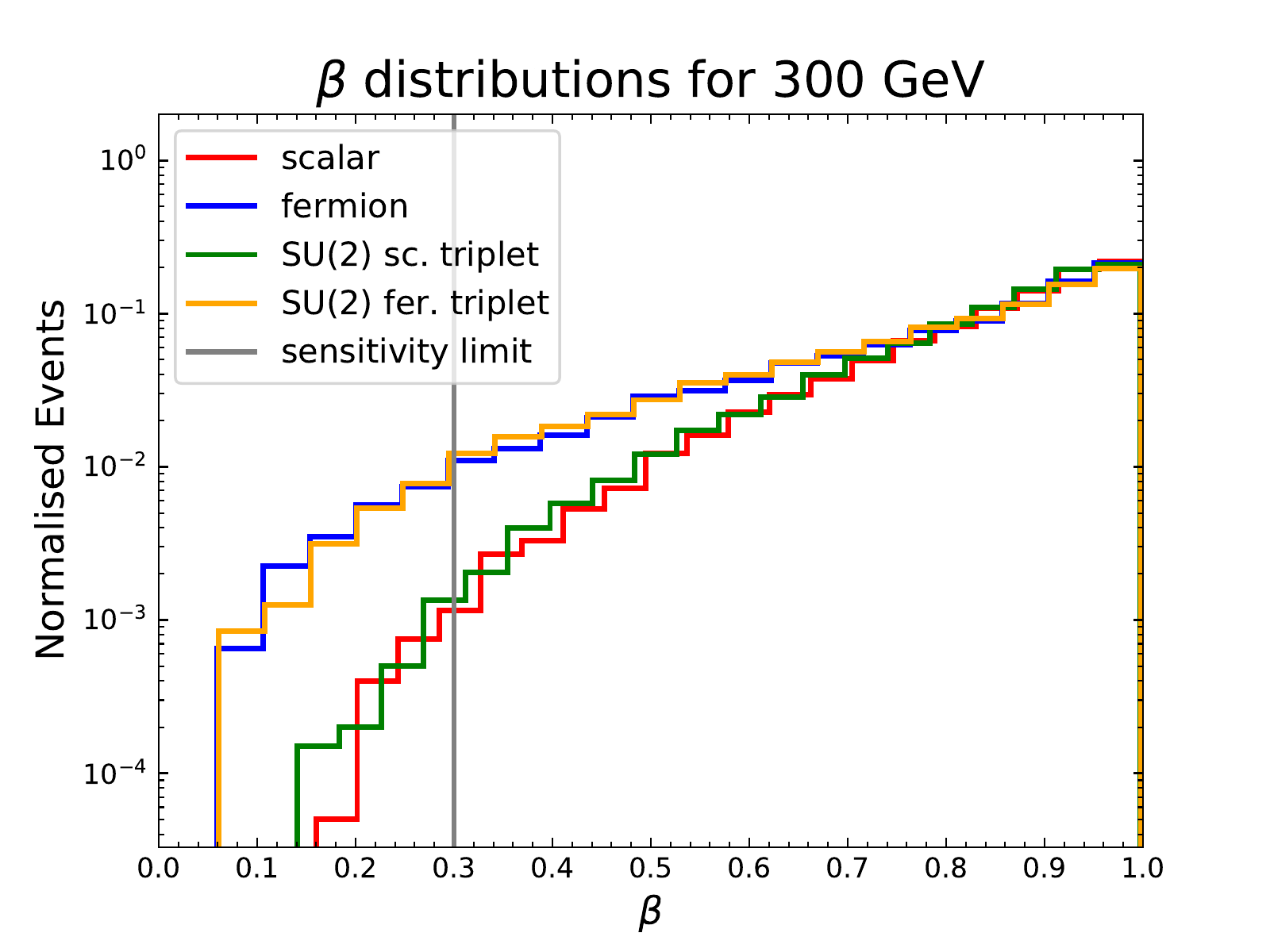}
\vspace{-1cm}
\caption{\it \small Top panel: the leading-order production cross-{s}ections
at 13 TeV for various types of doubly-charged particles. Middle and lower panels: 
the velocity distributions for various doubly-charged particles with
masses 50 and 300~GeV.}
\label{fig:2Q}
\end{figure}

We show in the top panel of Fig.~\ref{fig:2Q} the
13 TeV cross-{s}ections for the production of doubly-charged particles.
We see that, for species with the same spin,
the cross-sections for triplet particles are an order of magnitude larger 
than those for the singlet.
This is because we included all the neutral
$pp \to Y^0 Y^0, Y^+ Y^-, Y^{++} Y^{--} + \cdots$
and charged 
$pp \to Y^0 Y^\pm, Y^\pm Y^{\pm\pm} + \cdots$
triplet production modes,
assuming that the heavier components of the multiplets decay promptly into the nearly-degenerate 
lightest doubly-charged partner, so that all the production modes 
end up with the final state containing two doubly-charged particles.
The neutral charge combinations are produced via {$s$-channel $\gamma/Z$,
while the charged ones are produced via an $s$-channel $W^\pm$}.
Unlike the triplet species, the singlet species has only one neutral production mode,
namely $pp \to Y^{++} Y^{--} + \cdots$.

We also see in the top panel of Fig.~\ref{fig:2Q} that 
the cross-sections are an order of magnitude higher for fermions
than scalars with the same gauge quantum numbers.
One reason for this is that a Dirac fermion has twice as many degrees of freedom as a complex scalar. 
Another reason is, as already discussed in Section~\ref{sec:res_susy},
the production of doubly-charged particles is mediated by $s$-channel gauge boson exchange,
in which the two-scalar final states suffer threshold velocity suppression: 
$\sigma \to 0$ in the limit $\beta \to 0$.
This suppression is absent for fermion pair production.

In the case of doubly-charged particles, the threshold velocity for detection
is $\beta_{\rm th} = 0.3$, higher by a factor of two 
compared to singly-charged ones due to the higher electric charge.
In the middle and bottom panels of Fig.~\ref{fig:2Q} we show
the normalised velocity distributions 
for pair-produced doubly-charged particles  
with masses $m_{Y} = 50$ and 300 GeV, respectively.
The grey vertical lines indicate $\beta_{\rm th} = 0.3$, and only events
to the left of the lines are detectable by MoEDAL.
Comparing these two plots, we see that the distributions are more mass-dependent
in the high-$\beta$ region, whereas the mass effects are mild
in the lower-velocity regions where $\beta \lesssim 0.3$.
Nevertheless, about a factor of two more events satisfy $\beta < 0.3$ for $m_Y = 300$ GeV
than for $m_Y = 50$ GeV.
We also see clear differences in the momentum distributions between fermions and scalar particles 
for both $m_Y = 50$ and 300 GeV.
As discussed earlier, this is because angular-momentum conservation again
forces the scalar pair production via the $s$-channel gauge boson exchange
to be velocity suppressed, $\sigma \to 0$ in the limit $\beta \to 0$.
Unlike scalars, the production rate for fermions is non-vanishing even at $\beta = 0$.

\begin{figure}[]
\centering
\includegraphics[width=0.45\textwidth]{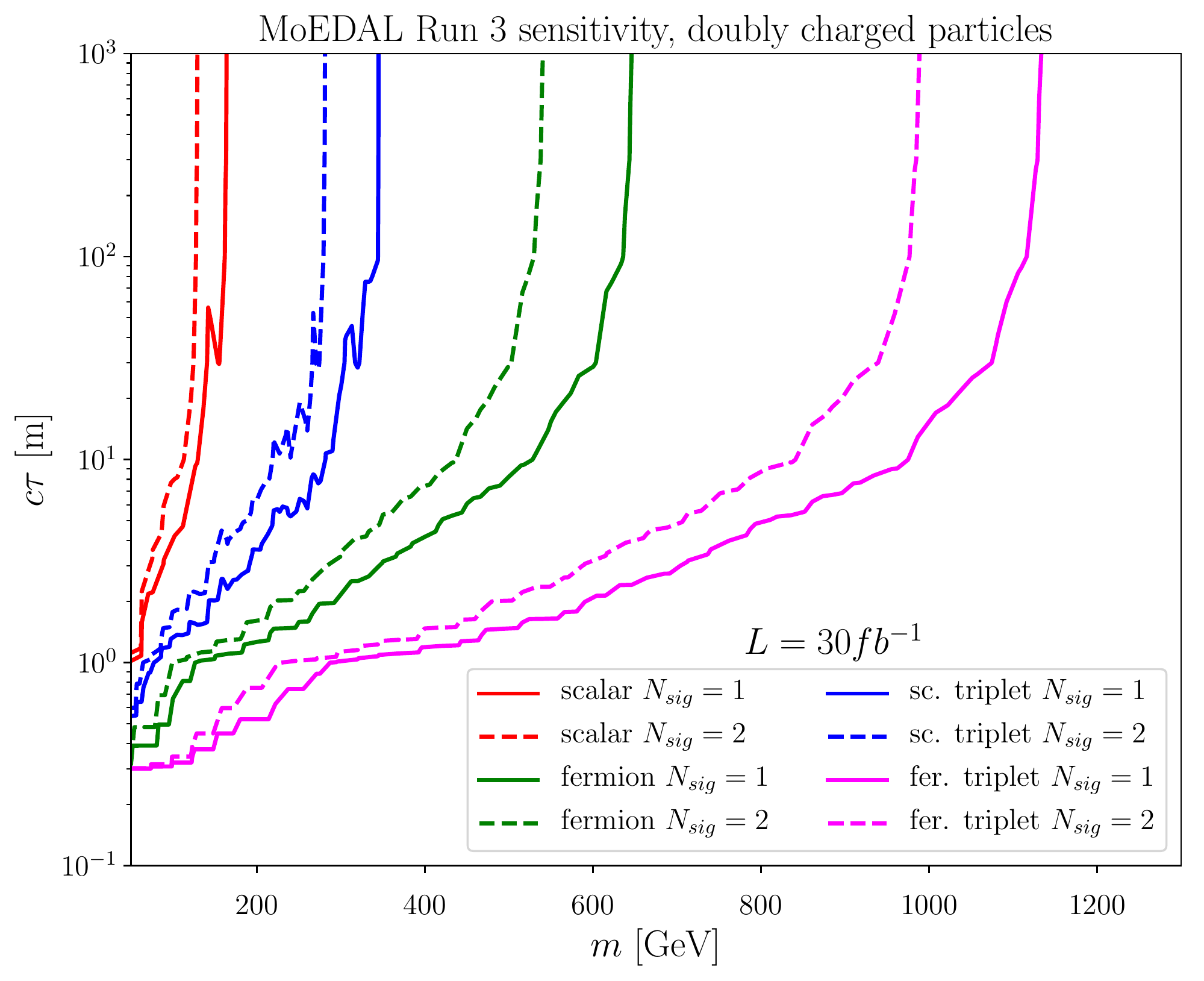}
\vspace{-0.5cm}
\caption{\it \small The expected sensitivities of MoEDAL for various long-lived doubly-charged particle species, assuming 30 fb$^{-1}$ of integrated luminosity.}
\label{fig:2Qsens}
\end{figure}

We show in Fig.~\ref{fig:2Qsens} the expected MoEDAL sensitivities for four types of
colour-singlet doubly-charged particles: a scalar singlet (red), a scalar triplet (blue), a fermion singlet (green)
and a fermion triplet (magenta), 
assuming a Run~3 integrated luminosity of 30 fb$^{-1}$.
As expected from the above discussion,
the fermion triplet has the highest mass reach
among the four types, due to its large cross-section and favourable production velocities. 
MoEDAL can probe this particle up to
{
$m_Y \sim
1130$ 
(990) 
GeV with $N_{\rm sig} = 1$ (2)}
for the Run~3 luminosity if $c \tau \gtrsim 100$\,m, as indicated by the solid (dashed) line.
The next most sensitive particle is the fermion-singlet, for which we estimate
the MoEDAL mass reach with $c \tau \gtrsim 100$\,m to be around { 
650 (540) GeV for $N_{\rm sig} = 1$ (2).}
The mass reaches for the two scalar particles (singlet and triplet) 
are significantly lower than the corresponding fermionic particles,
due to their smaller cross-sections and higher typical velocities.
The MoEDAL mass reaches at Run~3 are around 
{
340 (280)
}
GeV for the scalar triplet
and 
{
160 (130)
}
GeV for the scalar singlet for $N_{\rm sig} = 1$ (2) if $c \tau \gtrsim 100$\,m.

\subsection{Comparison with the existing searches}
\label{sec:comp_dcp}

We compare prospective MoEDAL mass reaches ($N_{\rm sig} = 1$) at Run~3
for the types of doubly-charged meta-stable particles 
examined in this Section, together with the available mass bound.
As discussed in Section~\ref{sec:comp_susy},
the most relevant constraints are obtained 
by the heavy stable charged particle (HSCP) searches
by ATLAS and CMS, which are based on
the ionization energy loss $dE/dx$ and 
the time-of-flight measurements but incorporate additional trigger requirements.
The analysis utilizing the largest data-set ($L = 36.1$ fb$^{-1}$)
is that from ATLAS~\cite{Aaboud:2019trc}.
In Section~\ref{sec:comp_susy} we compared
the MoEDAL mass reaches for various long-lived sparticles
with mass bounds from this analysis. 
However, ATLAS did not interpret their results for 
doubly-charged particles in~\cite{Aaboud:2019trc}, and did not provide mass limits.

\begin{table}[t!]
  \begin{center}
    \renewcommand{\arraystretch}{1.3}
  \begin{tabular}{|c||c|c|} 
\hline
& MoEDAL   & (CMS)  \\ 
\hline
Scalar singlet  & {160} & 
{((320))}\\ 
\hline
Fermion singlet  & {650} &
{(630)}\\
\hline
Scalar triplet  & {340} & 
{((590))} \\
\hline
Fermion triplet  & {1130} & {((900))}  \\
\hline
  \end{tabular}
\end{center}
\vspace{-2mm}
\caption{\it \small Compilation 
of the prospective MoEDAL mass reaches ($N_{\rm sig} = 1$)
at Run~3 and comparison with
the current ($95\%$ CL) mass bound 
for a long-lived fermion-singlet doubly-charged particle 
obtained from the CMS heavy stable charged particle search 
with $L = 2.5$ fb$^{-1}$~\cite{Khachatryan:2016sfv} (in parentheses). 
Our estimates of possible CMS bounds on other types of particle are in double parentheses. All masses are in GeV.
}
  \label{tab:dcp_limit}
\end{table}

CMS, on the other hand, has published their latest HSCP analysis \cite{Khachatryan:2016sfv}
based on a smaller data-set, $L = 2.5$ fb$^{-1}$, and
have interpreted their result for a long-lived doubly-charged 
particle that is an SU(2) singlet and has spin-1/2.
The mass bound for this particle is found to be 630 GeV,
which is 20 GeV lower than the MoEDAL Run~3 mass reach,
as shown in parentheses in Table~\ref{tab:dcp_limit}.
We emphasis that the CMS analysis incorporated extra assumptions, such as
an $E_{\rm T}^{\rm miss} > 170$~GeV requirement {or the presence of 
a muon-like particle}, rendering the result
more model-dependent than the prospective MoEDAL sensitivity.

In \cite{Khachatryan:2016sfv}, CMS did not directly interpret their result 
for other types of doubly-charged particles.
In order to derive approximate mass bounds 
for other types of doubly-charged particles, we assume 
that the detection efficiencies are not very sensitive to
the nature of particles other than the electric charge.
With this assumption we use the cross-section upper limit 
for the fermion singlet particle provided in~\cite{Khachatryan:2016sfv}
as approximate cross-section limits for 
other types of doubly-charged particles (scalar singlet, scalar triplet and fermion triplet).

We estimate the CMS mass bound for scalar singlet (scalar triplet)
obtained by the above recasting procedure to be {
$\sim 320 \, (590)$}~GeV, which can be compared with
the corresponding MoEDAL mass reach of {160 (340)} GeV
($N_{\rm sig} = 1$, $L = 30$ fb$^{-1}$).
In the case of the fermion triplet, we estimate the CMS mass bound to be {
$\sim 900$~GeV}
, which is surpassed by the Run~3 mass reach of MoEDAL, 
which is {1130} GeV.
Our estimates of these CMS sensitivities are shown in double parentheses, in view of {potential}
model dependence associated with the CMS analysis assumptions and the fact that they are
not official CMS results and have uncertainties due to our recasting procedure.

\section{Conclusions}
\label{sec:conx}

{We have analysed in this paper the {prospective}
sensitivities of the MoEDAL detector for searches
for singly- and doubly-charged long-lived particles during Run~3 of the LHC, considering
the specific examples of supersymmetric particles and scalars and fermions that are
suggested by Type-II and Type-III seesaw models of neutrino masses, respectively. We emphasise that the MoEDAL
search would be completely model-independent, using simply the capabilities of its NTDs.
However, MoEDAL suffers from two disadvantages with respect to the general-purpose LHC
detectors, ATLAS and CMS. It is sensitive only to an anomalous level of ionisation
corresponding to velocities $\beta < 0.15$ for singly-charged particles and $\beta < 0.3$
for doubly-charged particles, and it is anticipated that MoEDAL may accumulate 30~fb$^{-1}$
of luminosity by the end of Run~3, an order of magnitude less than ATLAS and CMS.}

{The greatest MoEDAL sensitivities for supersymmetric particles are for strongly-interacting
species, namely light-flavoured squarks, the gluino and stop squarks. There are also interesting
sensitivities for long-lived charginos, whereas the sensitivity for a long-lived slepton is
below the model-independent limit already established by LEP. {However}, 
we regard the model-independent potential MoEDAL sensitivities to other sparticles as quite complementary 
to the limits set by ATLAS and CMS, and of particular interest in scenarios such as 
those with weakly-broken R-parity in which there is no $E_{\rm T}^{\rm miss}$.}

{Among the particles examined in this paper,
the doubly-charged fermion triplet appears to be the most favourable for MoEDAL.
Since it is doubly-charged, the threshold velocity is relaxed 
to $\beta = 0.3$, so MoEDAL accepts a larger fraction of signal events.
Since it is a fermion, the particles are much more likely to be produced near threshold than
would be a scalar.
Finally, it is an SU(2)-triplet and has the largest cross section
among the doubly-charged particles we have studied.
Therefore, the typical mass scale that can be probed is higher than for the other electroweakly-interacting cases,
corresponding to a lower typical production velocity.
Although our study suggests that the MoEDAL can probe a substantial non-excluded 
region of the parameter space of the fermion-triplet doubly-charged particle, 
one must keep in mind that the estimated CMS limit is approximate 
and based on a much smaller data set than that obtained by the end of Run~2.}

It is clear that by the end of Run~3 ATLAS and CMS are likely to be able to provide 
better constraints on all the long-lived charged particle candidates that we have considered {here}.
However, we would like to emphasise that the MoEDAL analysis we have described here would
be able to set limits that are {independent of auxiliary signal assumptions.}

\section*{Acknowledgements}
The work of J.E.\ was supported by the UK STFC Grant ST/P000258/1 and by the Estonian Research Council via a Mobilitas Pluss grant.
D.K.G.\ would like to thank the High Energy, Cosmology and Astroparticle Physics
Division of ICTP, Trieste for hospitality during while part of this work was started. 
The work of R.M.\ is partially supported by
the National Science Centre, Poland, under research grants 2017/26/E/ST2/00135.
The work of K.S.\ is partially supported by the National Science Centre, Poland, under research grants 2017/26/E/ST2/00135 and
the Beethoven grants DEC-2016/23/G/ST2/04301.
D.K.G would like to thank N. Ghosh and I. Saha for discussions.

\end{document}